\documentclass[
 reprint,
superscriptaddress,
 amsmath,amssymb,
 aps,
pra,
longbibliography
]{revtex4-1}

\usepackage{graphicx}% Include figure files
\usepackage{dcolumn}% Align table columns on decimal point
\usepackage{bm}% bold math
\usepackage{bbm}% bold math
\usepackage[utf8]{inputenc}
\usepackage{amsfonts}
\usepackage{amsthm}
\usepackage{tikz}
\usepackage{physics}
\usepackage{comment}
\usepackage{xcolor}
\usepackage[caption=false,justification=justified]{subfig}
\usepackage{blindtext}
\usepackage{pgfplots,pgfplotstable}
\usepgfplotslibrary{polar}
\usepgfplotslibrary{colormaps}
\pgfplotsset{compat=newest}
\usepackage{hyperref}
\usepackage{xr-hyper}
\usepackage{footmisc}

\newcommand{\re}[1]{\Re[#1]}

\providecommand{\R}{\mathbb{R}}

\providecommand{\E}{\mathbb{E}}

\newcommand{\vac}{\ket{\mathrm{vac}}}
\newcommand{\cav}{\bra{\mathrm{vac}}}

\providecommand{\abs}[1]{\left\lvert#1\right\rvert}

\renewcommand{\d}{\mathrm{d}}
\newcommand{\e}{\mathrm{e}}
\newcommand{\ii}{\mathrm{i}}

\newcommand{\Var}[1]{\mathrm{Var}[#1]}

\newcommand{\acr}{\hat{a}^\dag}
\newcommand{\aan}{\hat{a}}
\newcommand{\bcr}{\hat{b}^\dag}
\newcommand{\ban}{\hat{b}}
\newcommand{\UBS}{U_{\mathrm{BS}}}
\newcommand{\hUBS}{\hat{U}_{\mathrm{BS}}}

\begin{document}

\title{Ultimate quantum sensitivity in the estimation of the delay between two interfering photons through frequency-resolving sampling}

\author{Danilo Triggiani}
\email{danilo.triggiani@port.ac.uk}
\affiliation{School of Mathematics and Physics, University of Portsmouth, Portsmouth PO1 3QL, UK}

\author{Giorgos Psaroudis}
\affiliation{School of Mathematics and Physics, University of Portsmouth, Portsmouth PO1 3QL, UK}

\author{Vincenzo Tamma}
\email{vincenzo.tamma@port.ac.uk}
\affiliation{School of Mathematics and Physics, University of Portsmouth, Portsmouth PO1 3QL, UK}
\affiliation{Institute of Cosmology and Gravitation, University of Portsmouth, Portsmouth PO1 3FX, UK}

\begin{abstract}
We demonstrate the ultimate sensitivity allowed by quantum physics in the estimation of the time delay between two photons by measuring their interference at a beam-splitter through frequency-resolving sampling measurements. This sensitivity can be increased quadratically by decreasing the photonic temporal bandwidth even at values smaller than the time delay when standard two-photon interferometers become inoperable and without adapting the path of the reference photon, nor the need of time-resolving detectors with an unfeasible high resolution. Applications can range from more feasible imaging of nanostructures, including biological samples, and nanomaterial surfaces to quantum enhanced estimation based on frequency-resolved boson sampling in optical networks.
\end{abstract}

\pacs{Valid PACS appear here}

\maketitle

Photons manifest unique quantum properties which might appear odd and counter intuitive from a classical standpoint. 
A paradigmatic example of this unusual behaviour is given by the interference of two identical photons after impinging separately at the two faces of a balanced beam-splitter, as the two photons always end up together in one of the two output arms of the beam-splitter~\cite{Hong1987,Shih1988,Abram1986,Prasad1987,Ou1987, Fearn1987, Rarity1989,Bouchard2021}. 
This tendency of the photons to `bunch' together arises from the lack of information on which path the two photons undertake when interfering at the beam-splitter. 
Such interference phenomenon is sensitive to the differences between the physical parameters associated with the two photons.
This has motivated estimation schemes for high-precision measurements of the two-photon time delay~\cite{Hong1987,Lyons2018,Chen2019,Fabre2021}, or the state of polarization of the photons~\cite{Harnchaiwat2020}.
Remarkably, the estimation of time intervals has achieved in this way precisions ranging from sub-picoseconds~\cite{Hong1987} up to the attoseconds regime~\cite{Lyons2018}. 
In particular, the employment of notions borrowed from estimation theory and the subsequent study of the Fisher information~\cite{Cramer1999,Rohatgi2000} -- a way to quantify the ultimate amount of information that can be obtained about an unknown parameter with a given estimation scheme -- have recently allowed for a further boost in the levels of precision achievable~\cite{Lyons2018,Chen2019,Scott2020,Fabre2021}.
On the other hand, these analyses also show that the sensitivity of schemes based purely on the observation of the coincidence and bunching statistic is highly dependent on how much the two photons differ from each other in the parameter to estimate, such as their relative time delay~\cite{Hong1987, Shih1988, Abram1986, Prasad1987, Ou1987, Fearn1987, Rarity1989, Lyons2018, Bouchard2021}. 
In particular, such two-photon interference techniques become insensitive to photonic time delays beyond the temporal bandwidth of the photons~\cite{Hong1987, Shih1988, Abram1986, Prasad1987, Ou1987, Fearn1987, Rarity1989, Lyons2018, Bouchard2021}

A different approach with respect to standard two-photon interference takes advantage of current detectors capable to resolve inner-mode variables of the photons, such as their time of arrival~\cite{Legero2003,Legero2004,Tamma2015,Wang2018, Scott2020,Prakash2021}, or their frequencies~\cite{Jin2015,Yepiz-Graciano2020,Hiemstra2020}. 
Indeed, one can exploit the quantum beating, i.e. the oscillations in the count of coincidence and bunching events of two photons as a function of the detection times or the detected frequencies, to infer the value of the frequency-shift or the time delay of the two photons, respectively~\cite{Legero2003,Legero2004,Laibacher2018,Orre2019}. 
Furthermore, differently from non-resolved two-photon interference (i.e. standard interferometry performed without resolving the photonic inner-mode variables) quantum beating can be observed even in the case of negligible overlap in the time or frequency domain between the photonic wave-packets impinging at the beam-splitter. 
However, although a quantum advantage for inner-mode variables interference of independent photons has been already demonstrated from a computational point of view~\cite{Laibacher2015, Tamma2016, Tamma2021}, the ultimate precision of such technique from a metrological perspective is yet not known. 
This motivates the following important questions pertinent to the development of high-precision quantum sensing technologies, such as feasible observation and imaging of nanomaterials or nanostructures in biological samples~\cite{Lyons2018}.
Is there a quantum metrological advantage arising from the inner-mode variables quantum interference of photons? 
If yes, is it possible to quantify such an advantage in terms of the ultimate precision fundamentally achievable? 
How does such an advantage depend on the value of the parameter to estimate?

\begin{figure}[t!]
\centering
\includegraphics[width=.95\columnwidth]{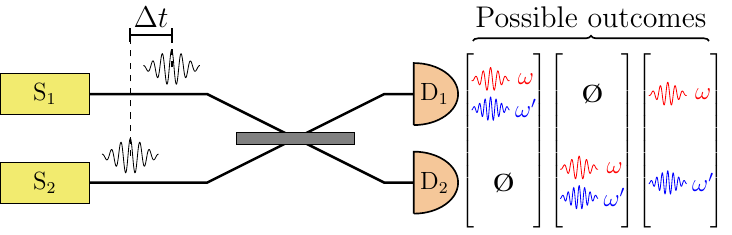}
\caption{Scheme of the interferometric setup. Two independent photons in the state in Eq.~\eqref{eq:InitState} with identical frequency distributions are produced by independent sources $\mathrm{S}_1$ and $\mathrm{S}_2$ with a given degree of indistinguishability $\eta$ deriving from non-temporal properties (e.g. different polarizations).
After impinging onto the two faces of a balanced beam-splitter with an unknown time delay $\Delta t$, the two photons are eventually observed by the frequency-resolving 
detectors $\mathrm{D}_1$ and $\mathrm{D}_2$. 
At each experimental run, a sample $(\omega, \omega', \mathrm{X})$ from the output probability distribution in Eq. ~\eqref{eq:P} is observed, where $\omega$ and $\omega'$ are the random values of the frequencies of the detected photons and X$\,$=$\,$B,C specifies if it is a bunching (B) or a coincidence (C) event.
}
\label{fig:Setup}
\end{figure}

In this work, we demonstrate the ultimate precision achievable in the estimation of time delays between independent photons with same frequency distribution, interfering at a balanced beam-splitter, and detected at the output with frequency-resolving detectors, as described in \figurename~\ref{fig:Setup}. 
We show that the measurement scheme proposed is optimal for identical photons with an arbitrary time delay.
When instead a non-vanishing distinguishability between the two photons in any non-temporal property (e.g. polarizations) is present, the quantum advantage is still retained, even if such non-temporal distinguishability is not `erased' at the detectors by suitable measurements.
In particular, we demonstrate that this scheme is effective also in the regime of photonic temporal bandwidth much smaller than their temporal delay, a regime in which non-resolving two-photon interferometry fails.
This is a particularly interesting feature since, as we will show, it allows us to employ extremely short photons, which have the largest sensitivity in the estimation of time delays, without the need of fine-tuning the reference path of the interferometer with a precision comparable with the temporal bandwidth of the photons.
We finally perform a numerical simulation showing how the ultimate precision assessed by the Fisher information we evaluate can be practically achieved with a feasible number of iterations of the experiment.

\section{Experimental setup}
The quantum state of the two independent photons before impinging on the beam-splitter is (see \figurename~\ref{fig:Setup})

\begin{align}
\ket{\psi}=\int_{\R} &\d\omega_1\ \xi_1(\omega_1)\left(\eta\acr_{1,\omega_1}+\sqrt{1-\eta^2}\bcr_{1,\omega_1}\right)\vac \notag\\
&\ \otimes \int_{\R}d\omega_2\ \xi_2(\omega_2)\acr_{2,\omega_2}\vac,
\label{eq:InitState}
\end{align} 
where $\xi_i(\omega)= \bar{\xi}(\omega)\e^{-\ii\omega t_i}$ is the frequency probability amplitude of the photon injected in the $i$-th channel, with $i=1,2$, with $\abs{\xi_1(\omega)}=\abs{\xi_2(\omega)} \equiv \bar{\xi}(\omega)$, (photons with the same spectra) and $t_i$ is the time of incidence of the $i$-th photon on the beam-splitter. 
The quantity $\Delta t=t_2-t_1$ is thus the unknown delay to be measured.
For a given frequency $\omega$ and spatial channel $i$, the commuting creation operators $\hat{a}^\dag_{i,\omega}$ and $\hat{b}^\dag_{i,\omega}$ denote orthogonal modes in any given additional degree of freedom, such as polarisation or inner spatial modes: for $\eta=1$, the two photons differ only in their injection times, while for $0\leqslant\eta<1$ a further distinction between the states of the two photons arises (e.g. they can be in different polarisations, where $\eta = 0$ is the limit case of orthogonal polarizations).

In our model, at each repetition of the experiment, the photons are randomly observed at either of the output ports of the balanced beam-splitter.
Simultaneously, their frequencies are measured at all possible random values within their frequency distribution $\abs{\xi(\omega)}^2$ (with a given spectral bandwidth $\sigma$), with high enough resolution $\delta\omega$ so that their distinguishability in time is erased, i.e., for Gaussian spectra,~\cite{Jin2015}
\begin{equation}
\delta\omega \ll \frac{1}{\abs{\Delta t}} ,\quad\mathrm{and}\quad \delta\omega\ll\sigma.
\label{eq:Resolution}
\end{equation}
Noticeably, the second condition on $\delta \omega$ in Eq.~\eqref{eq:Resolution} is weaker than the former in the regime of highest precision, which we will show to be for large $\sigma$.
For imperfect detectors able to detect an incoming photon with a finite probability $\gamma<1$,
\begin{multline}
P_\eta(\omega,\omega',\mathrm{X})=\gamma^2\bar{\xi}(\omega)^2\bar{\xi}(\omega')^2\\ 
\times\left(1+\alpha(\mathrm{X})\eta^2\cos((\omega-\omega')\Delta t)\right),
\label{eq:P}
\end{multline}
with X$\,$=$\,$B,C, $\alpha(B)=1$ and $\alpha(C)=-1$, represents the probability of the sample outcome $(\omega,\omega',\mathrm{X})$ of a single iteration of the experiment, in which the two photons either bunch in the same channel (X$\,$=$\,$B) or end up in different channels (X$\,$=$\,$C) with random frequency values $\omega$ and $\omega'$, up to a factor $\delta\omega^2$ due to the resolution of the detectors (see Appendix~\ref{app:ProbCal}).
Expectedly, the probability in Eq.~\eqref{eq:P} manifests quantum beats with period inversely proportional to the photonic time delay $\Delta t$.

The experiment thus consists in sampling from the output probability distribution in Eq.~\eqref{eq:P} the outcomes $(\omega,\omega',\mathrm{X})$, without spectral filtering at specific frequencies (see Appendix~\ref{app:Num}). 
Therefore, similarly to boson sampling~\cite{Laibacher2015, Tamma2016, Tamma2021}, the output probability distribution does not need to be experimentally reproduced avoiding therefore the need of a large number of experimental runs. 
Indeed, we will show that, for photons with Gaussian frequency distribution, it is enough to observe $<1000$ samples, and thus photonic pairs, to achieve unbiasedness and optimal precision (see Appendix~\ref{app:Num}).

\section{Bounds on the precision}

To fully analyse the precision achievable with our setup in \figurename~\ref{fig:Setup}, we evaluate the bound on the variance $\Var{\widetilde{\Delta t}}$ of any unbiased estimator $\widetilde{\Delta t}$ associated with our scheme, given by the Cramér–Rao bound~\cite{Cramer1999,Rohatgi2000}, and compare it with the smallest variance achievable with any measurement scheme, given by the quantum Cramér–Rao bound~\cite{Helstrom1969,Holevo2011}.
These bounds are related by the chain of inequalities
\begin{equation}
\Var{\widetilde{\Delta t}}\geqslant\frac{1}{N F(\Delta t)} \geqslant \frac{1}{N H(\Delta t)},
\label{eq:CRB}
\end{equation}
where $N$ is the number of repetitions of the measurement, $F(\Delta t)$ is the Fisher information associated with the frequency-resolving estimation scheme~\cite{Cramer1999,Rohatgi2000}, and $H(\Delta t)$ is the quantum Fisher information, i.e. the maximum of the Fisher information over all possible measurement schemes employing the photonic state in Eq.~\eqref{eq:InitState}~\cite{Helstrom1969,Holevo2011}.

The first inequality in Eq.~\eqref{eq:CRB} can always be saturated in the asymptotic regime of large $N$ by the maximum-likelihood estimator~\cite{Cramer1999,Rohatgi2000} (see Appendix~\ref{app:Num}), so we will focus on the analysis of $F(\Delta t)$.
The maximum precision achievable with the probe state in Eq.~\eqref{eq:InitState}, given by the quantum Fisher information  (see Appendix~\ref{app:QFI})
\begin{equation}
H(\Delta t) = 2\sigma^2 = \frac{1}{2\tau^2} \equiv H,
\label{eq:QFI}
\end{equation}
is independent of the value of $\Delta t$ to be estimated,
with $\sigma^2$ the squared spectral bandwidth of each photon frequency probability distribution, i.e. the variance of $\bar{\xi}(\omega)^2$, and $\tau=1/2\sigma$ the temporal bandwidth. 
Compared to the quantum Fisher information in the case of entangled photons generated with spontaneous parametric down-conversion with spectral bandwidth $\sigma$, the one in Eq.~\eqref{eq:QFI} is halved~\cite{Chen2019,Scott2020,Jordan2022}.
This means that, by only relying on independent photons, it is possible to achieve an ultimate precision which differs only by a constant factor $1/\sqrt{2}$ from the one achievable with entangled photons. 

\section{Fisher information based on frequency-resolved measurements}

We now determine the expression of the Fisher information for frequency-resolved measurements in the setup in \figurename~\ref{fig:Setup}. 
This includes the contribution from all the possible frequency-resolved events of photon bunching and coincidences occurring with the probabilities in Eq.~\eqref{eq:P}, respectively. 
The Fisher information for such scheme becomes (see Appendix~\ref{app:FI})
\begin{align}
F_\eta(\Delta t)=\eta^4\gamma^2\mathcal{I}_\eta(\Delta  t),
\label{eq:FisherGen}
\end{align}
with
\begin{equation}
\mathcal{I}_\eta(\Delta  t) = \int_{\R^2}\!\! \d\omega\d\omega'\ \! \bar{\xi}(\omega)^2\bar{\xi}(\omega')^2 (\omega-\omega')^2\zeta_\eta((\omega-\omega')\Delta t),
\label{eq:Integral}
\end{equation}
where 
\begin{equation}
\zeta_\eta(x)=\frac{\sin^2 x}{1-\eta^4\cos^2 x}
\label{eq:zeta}
\end{equation}
is for $\eta\neq 1$ a periodic function of period $\pi$ oscillating between $0$ and $1$, whilst for $\eta=1$ it becomes identically equal to $1$.

We consider first the case where the photons differ only in the time they impinge on the beam-splitter ($\eta =1$). Due to the photon indistinguishability at the detectors arising from the frequency-resolved measurement, the Fisher information $F_{\eta=1}(\Delta t) \equiv F_{\eta=1}$ becomes independent of $\Delta t$ and proportional to the quantum Fisher information in Eq.~\eqref{eq:QFI}, i.e. 
\begin{equation}
F_{\eta=1} = \gamma^2 H = 2\gamma^2\sigma^2= \frac{\gamma^2}{2\tau^{2}},
\label{eq:FisherEta1}
\end{equation}
implying that shorter photons yield a more precise estimation.
Remarkably, the detector efficiency only affects the estimation precision through a constant factor $\gamma^2$, and, for lossless detectors $\gamma=1$, the Fisher information equals the quantum Fisher information $H$ found in Eq.~\eqref{eq:QFI}.
Therefore, this estimation scheme based on frequency-resolving measurements is optimal, and the level of high precision achieved is independent of the value of the delay to be estimated.
Furthermore, $F_{\eta=1}$, similarly to $H$, is independent of the structure of the photonic wave-packets, and it increases with the spectral variance $\sigma^2$, for any value of the time delay $\Delta t$.
In practical terms, if we assume a biphoton rate of 1 MHz and lossless detectors, employing photons with a temporal bandwidth of $\tau\sim$ 60 fs within the reach of the state of the art~\cite{Mosley2008,Nasr2008}, our frequency-resolving technique allows us to reach the attosecond precision in the estimation of any delay in only 2 hours of measurements, while reducing $\tau$ to 10 fs would allow us to achieve the same sensitivity in less than 4 minutes.
In addition to high-precision measurements, our technique also allows for faster estimations.

We consider now the case $\eta<1$ where the photons manifest some distinguishability at the detectors in parameters other than time. 
Of particular interest is the regime of small values of $\tau/\Delta t$. 
Indeed, when shorter and shorter temporal bandwidths $\tau$, even smaller than $\Delta t$, are employed, the quantum Fisher information in Eq.~\eqref{eq:QFI}, yields higher and higher sensitivities.
The value of the Fisher information in Eq.~\eqref{eq:FisherGen} in such regime, for regular photonic spectra (see Appendix~\ref{app:FILargeDelay})
\begin{align}
F_\eta(\tau/\Delta t\ll 1) = (1-\sqrt{1-\eta^4})F_{\eta=1}\propto\frac{1}{\tau^2}
\label{eq:FisherGenBig}
\end{align}
is, for any value of $\eta <1$, independent of $\Delta t$, proportional to the spectral variance $\sigma^2$ and independent of any other parameter in the wave-packet distribution, as the quantum Fisher information $H$ in Eq.~\eqref{eq:QFI}.

The Cramér–Rao bound in Eq.~\eqref{eq:CRB} associated with the Fisher information in Eq.~\eqref{eq:FisherEta1} and Eq.~\eqref{eq:FisherGenBig} entails, for any value of $\eta$, an increase of precision when broader spectral bandwidths $\sigma$ (shorter temporal bandwidths $\tau$) are employed.
This also overcomes the need of optimising the reference path of the interferometer with high precision of the order of $\tau$ to maximise the Fisher information as customary for non-resolved two-photon interference~\cite{Lyons2018}.

\section{Example of Gaussian wave-packets}
\label{sec:Gaussian}

A typical experimental example is the case of Gaussian wave-packets for which the Fisher information in Eq.~\eqref{eq:FisherGen} reduces to (see Appendix~\ref{app:FIGauss})
\begin{align}
F_\eta^\mathrm{G}(\Delta t)=F_{\eta=1}\,\mathcal{I}^\mathrm{G}_\eta\left(\frac{\Delta t}{\tau}\right),
\label{eq:FisherGauss}
\end{align}
with the integral
\begin{equation}
\mathcal{I}^\mathrm{G}_\eta\left(\frac{\Delta t}{\tau}\right) =2\eta^4\sqrt{\frac{1}{\pi}} \int_{-\infty}^\infty\d \kappa\ \e^{-\kappa^2}\kappa^2 \zeta_\eta\left(\kappa\frac{\Delta t}{\tau}\right)
\label{eq:GaussIntegral}
\end{equation}
dependent on the periodic function $\zeta_\eta$ in Eq.\eqref{eq:zeta}.

\begin{figure}[t!]
\centering
\subfloat{\includegraphics[width=.975\columnwidth]{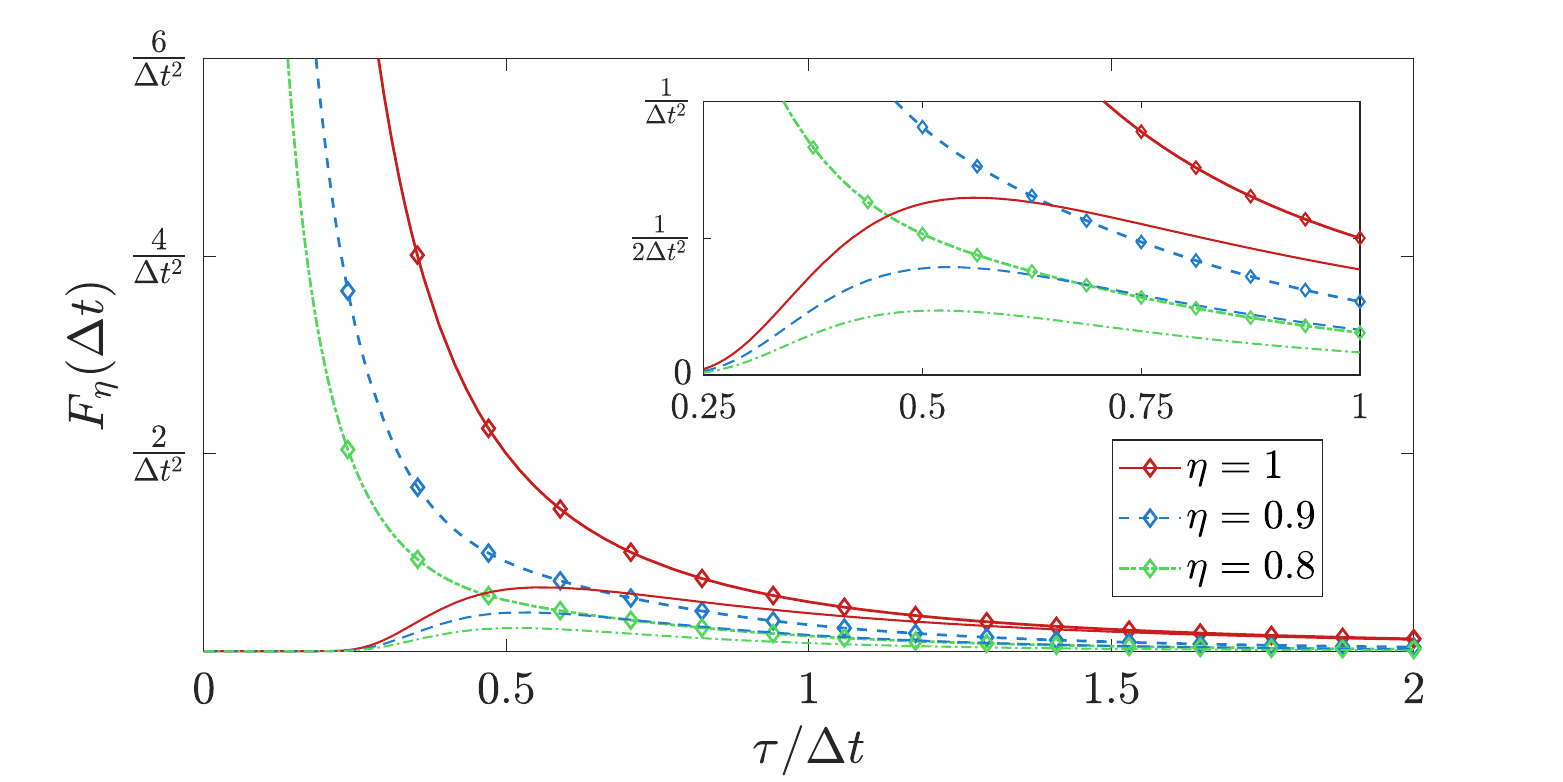}}
\caption{Plots of the Fisher information $F_\eta(\Delta t)=F_\eta^{\mathrm{G}}(\Delta t)$ for Gaussian photons in Eq.~\eqref{eq:FisherGauss} (lines with markers) and $F_\eta(\Delta t)=F_\eta^{\mathrm{G,NR}}(\Delta t)$ in Eq.~\eqref{eq:FisherGaussNoFreq} (thinner lines) as functions of the ratio $\tau/\Delta t$ between the photonic temporal bandwidth and the delay to be estimated.
The non-resolving Fisher information quickly decreases for temporal bandwidths smaller than the optimal value $\tau \simeq \Delta t /2$, while the frequency-resolving Fisher information rapidly increases (see Eq.~\eqref{eq:FisherGenBig}).
This means that, for any given value of the delay, the quantum metrological advantage of resolved measurements increases in principle arbitrarily when employing shorter and shorter photons.
}
\label{fig:FisherGauss}
\end{figure}

As a practical comparison, the Fisher information in Eq.~\eqref{eq:FisherGauss} manifests a clear metrological advantage with respect to the Fisher information associated with non-resolved (NR) two-photon interference measurements (see Appendix~\ref{app:FINR})
\begin{equation}
F_\eta^{\mathrm{G,NR}}(\Delta t) =  F_{\eta=1}\, \frac{1}{2} \frac{\eta^4}{\exp[\frac{\Delta t^2}{2\tau^2}]-\eta^4}\frac{\Delta t^2}{\tau^2}.
\label{eq:FisherGaussNoFreq}
\end{equation}
In fact, as evident from \figurename~\ref{fig:FisherGauss}, $F^\mathrm{G}_\eta(\Delta t) > F^\mathrm{G,NR}_\eta(\Delta t)$ independently of the indistinguishability $\eta$ at the detectors, of the photonic spectral bandwidth $\sigma$, and of the value $\Delta t$ to be estimated, leading to a quantum advantage of frequency-resolved measurements. 
In particular, for any given value of the delay $\Delta t$, when shorter and shorter photons are employed to maximize the quantum Fisher information, such quantum advantage $F^\mathrm{G}(\Delta t)/F^\mathrm{G,NR}(\Delta t)$ increases exponentially with $1/\tau$ (see Appendix~\ref{app:FIGauss}).
Indeed, in absence of information about the frequencies of the detected photons in the non-resolved technique, the estimation has to be restricted only to feasible experimental scenarios $\Delta t\simeq 2\tau$ (see \figurename~\ref{fig:FisherGauss}), and it is not possible to take full advantage of the quadratic scaling with $1/\tau$ of the quantum Fisher information.

\begin{figure*}[t!]
\centering
\begin{minipage}{.45\linewidth}
\subfloat{\includegraphics[width=\linewidth]{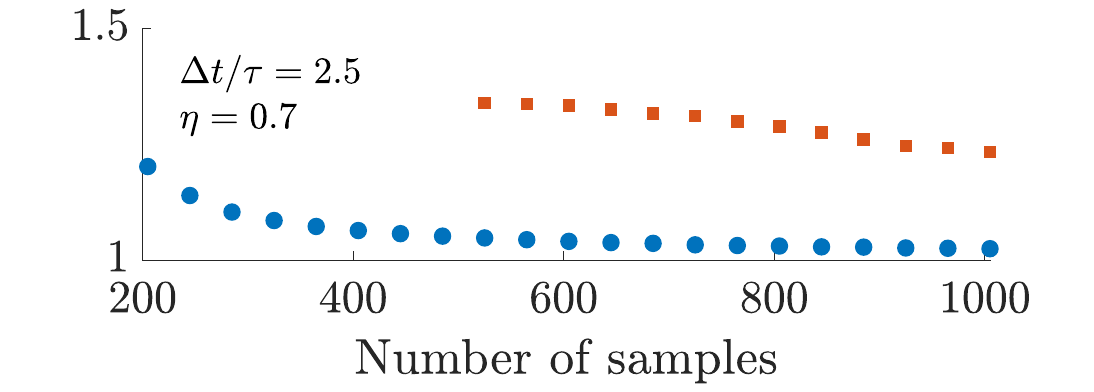}}\vspace{-12pt}\\
\subfloat{\includegraphics[width=\linewidth]{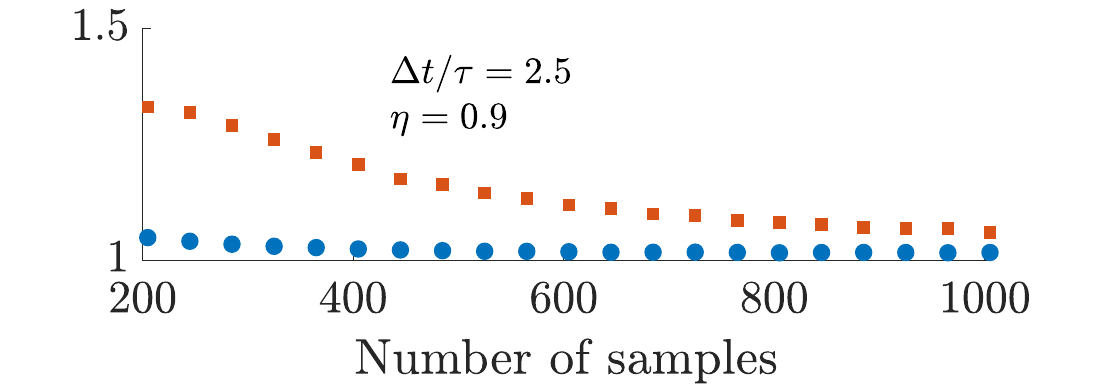}}
\end{minipage}\hspace{-10pt}
\begin{minipage}{.45\linewidth}
\subfloat{\includegraphics[width=\linewidth]{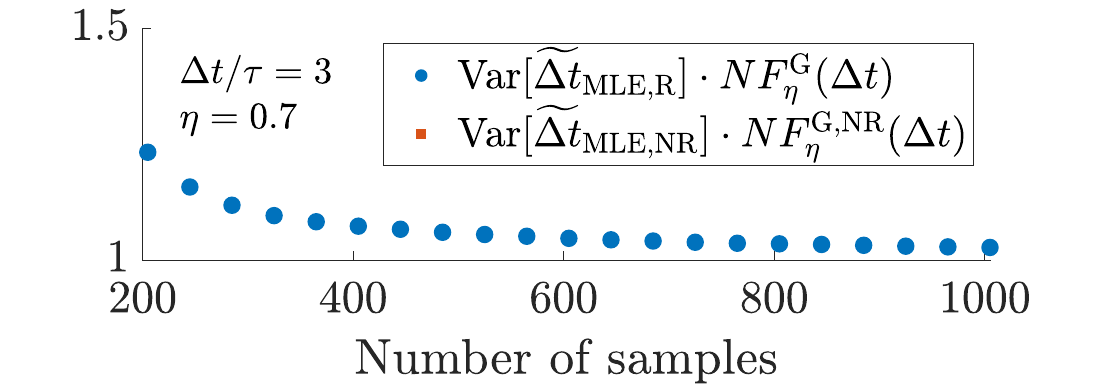}}\vspace{-12pt}\\
\subfloat{\includegraphics[width=\linewidth]{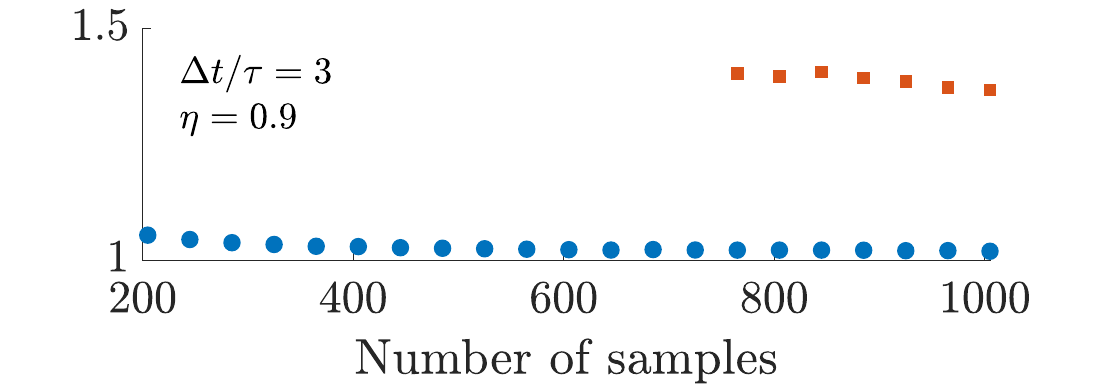}}
\end{minipage}
\caption{Four numerical simulations have been carried out to test the rate of convergence of the variance of the maximum-likelihood estimator to the correspondent Cramér–Rao bound in Eq.~\eqref{eq:CRB} for the resolving (blue circles) and non-resolving (orange squares) technique for Gaussian spectra. 
Each point represents an average over $10^6$ attempted iterations of the estimation. 
For the non-resolved approach, points where more than $1\%$ of the estimations failed to yield a finite estimate were omitted (see Appendix~\ref{app:Num}).
Our technique appears to saturate the Cramér-Rao bound with fewer samples than in the non-resolving approach, particularly for decreasing values of $\tau/\Delta t$ and decreasing $\eta$.
}
\label{fig:MLENumeric}
\end{figure*}

As a further advantage of the frequency-resolving technique, from the results obtained with numerical simulations shown in \figurename~\ref{fig:MLENumeric}, it appears that the saturation of the Cramér-Rao bound in Eq.~\eqref{eq:CRB} when employing the maximum-likelihood estimator~\cite{Cramer1999,Rohatgi2000} is consistently achieved with less than $1000$ observed samples independently of the value of the delay, differently from the non-resolving approach (see Appendix~\ref{app:Num}).
Moreover, the outcome of the frequency-resolving estimation appears to always yield finite estimates of $\Delta t$. Instead, in the non-resolving approach, the observed number of bunching and coincidence events can fail to provide a real and finite estimate of $\Delta t$ due to statistical fluctuations.
The corresponding fail probability drastically increases for larger delays, even if still of the order of $\tau$, and for smaller values of the distinguishability $\eta$, even when we restrict ourselves to optimal values of the delay (see Appendix~\ref{app:Num}).

\section{Conclusions}

We have demonstrated quantum metrological advantage  in the estimation of the time delay between two photons with same frequency distribution in a two-photon interference setup with frequency-resolving detectors. 
We have shown that, for photons with arbitrary time delay, sampling from the coincidence and bunching probabilities, and additionally from all the possible photonic frequencies, is an optimal estimation strategy -- in the sense that it saturates the ultimate precision given by the quantum Cramér–Rao bound.
 
Even for pairs of photons with any degree of non-temporal distinguishability in the measurement (e.g. polarization), this frequency-resolved technique still outperforms non-resolving measurements, for any value of the spectral bandwidth and of the time delay to be estimated.
This enhancement in precision is especially significant with ultra-short photons, even shorter than the time delay to be estimated, since both the quantum Fisher information and the frequency-resolving Fisher information increase quadratically with the photonic spectral bandwidth.
Indeed, independently of the value of the time delay to estimate, there is no upper limit, in principle, to the maximum achievable precision, particularly in view of future sources able to generate shorter and shorter photons. 
This is not the case of non-resolved interferometers, where the maximum precision (achievable only after a prior optimisation of the setup) is fundamentally limited in value for any fixed delay, and the sensitivity decreases exponentially with smaller and smaller overlap between the photonic wave-packets~\cite{Lyons2018}.
Moreover, it appears that, compared to the non-resolving technique, fewer samples are required to saturate the Cramér-Rao bound through the maximum-likelihood estimator.

Furthermore, the precision of this technique is not limited by the detector time resolution required to resolve directly the time delay as in time-resolved measurements~\cite{Scott2020}. 
Indeed, this constraint is circumvented by a much less demanding resolution in the frequency domain which aims only to ensure photon indistinguishability in time at the detectors independently of the precision one wants to achieve in the estimation. 
Instead, such a precision is fundamentally limited mainly by the minimum achievable temporal bandwidth of the photons. 
These results have therefore the potential to substantially improve the precision and the duration of the measurement of current and future sensing techniques (e.g. optical coherence tomography~\cite{Aumann2019}, quantitative phase microscopy~\cite{Nguyen2022}, holographic interference microscopy~\cite{Tishko2011}) employing photons that are shorter than the delay to be measured, with applications, among others, in biomedical sensing~\cite{Drexler2001,Park2018}, such as surface roughness characterization~\cite{Saarakkala2009}, and more in general in the characterization of any semi-transparent sample.

From a more fundamental perspective, the  proposed technique unravels the full metrological power of frequency-resolved sampling measurements allowing to infer high-precision information about the photon in the conjugate (time) domain that would otherwise be completely lost. 
We also emphasize that it is possible to operate the same technique to enhance the precision of two-photon interference protocols for the estimation of the frequency shift between two independent photons through time-resolving measurements, i.e. inverting the role of the two conjugate variables.
These results are thus applicable, in principle, to general estimation schemes based on any pair of conjugate parameters, such as position and momentum or angular position and orbital angular momentum. 
They also lay the foundations for future schemes based on multi-parameter resolved measurements able to increase the indistinguishability of photons at the detectors in multiple degrees of freedom, and therefore enhance the metrological capabilities of multi-photon interference techniques.

Finally, an extension of such results to more than two photons could lead in future works to the demonstration of a quantum metrological advantage of inner-mode variables multi-photon interference in more general linear optical networks with a larger number $N$ of non-entangled input photons based on sampling in suitable photonic inner degrees of freedom at the interferometer output~\cite{Laibacher2018}.
The extension of this technique to optical network with two or more entangled photons will be also addressed in future works.

\section*{Acknowledgements}
We thank Paolo Facchi and Frank A. Narducci for the helpful discussions.
This work was partially supported by the Office of Naval Research
Global (N62909-18-1-2153).

\appendix

\section{Quantum Fisher information}
\label{app:QFI}
In this section we will evaluate the Quantum Fisher information in Eq.~\eqref{eq:QFI} in the main text.
To do so, we first need to rewrite the probe state $\ket{\psi}$ in Eq.~\eqref{eq:InitState} in terms of the delay $\Delta t$ between the two photons. In the following treatment, we will suppose that all the moments of the frequency distribution of the two photons are known, as well as the distinguishability parameter $\eta$, while the time delay $\Delta t = t_1 - t_2$ and the sum of the emission times $t_\mathrm{tot}=t_1+t_2$ are unknown, as it is customary in experimental scenarios. Since in this setup two unknown parameter are present, the formal approach to this estimation problem requires the employment of the $2\times 2$ Quantum Fisher information matrix (QFIM) $\mathcal{H}$ \cite{Helstrom1969,Holevo2011,Paris2009}, whose elements are given by
\begin{equation}
\mathcal{H}_{ij} = 4\Re[\braket{\partial_i \psi}{\partial_j \psi}-\braket{\psi}{\partial_i \psi}^*\braket{\psi}{\partial_j \psi}],\quad i,j=1,2,
\label{eq:QFIM}
\end{equation}
where we denoted with $\partial_{1} \equiv \partial_-$ and $\partial_2\equiv \partial_+$ the derivatives with respect to $\Delta t$ and $t_\mathrm{tot}$ respectively, while the Quantum Cramér-Rao bound is given in matrix form
\begin{equation}
\mathrm{Cov}[\{\Delta t, t_{\mathrm{tot}}\}] \geq \frac{\mathcal{H}^{-1}}{N}.
\end{equation}
We will show that the QFIM is diagonal, so that the Cramér-Rao bound associated with the time delay can be written only in terms of $\mathcal{H}_{11} \equiv H(\Delta t)$ with $H(\Delta t)$ shown in Eq.~\eqref{eq:QFI}.

First, we rewrite the state $\ket{\psi}$ from Eq.~\eqref{eq:InitState}

\begingroup
\allowdisplaybreaks
\begin{align}
\ket{\psi} &= \int_{\R^2} \d\omega_1\d\omega_2\ \bar{\xi}(\omega_1)\bar{\xi}(\omega_2)\e^{-\ii \omega_1 t_1 -\ii \omega_2 t_2}\notag\\
&\quad\times\left(\eta\acr_{1,\omega_1}+\sqrt{1-\eta^2}\bcr_{1,\omega_1}\right)\acr_{2,\omega_2}\vac,\notag\\
&=\int_{\R^2} \d\omega_1\d\omega_2\ \bar{\xi}(\omega_1)\bar{\xi}(\omega_2)\e^{-\ii t_{\mathrm{tot}}(\omega_1+\omega_2)/2 - \ii \Delta t(\omega_1-\omega_2)/2}\notag\\
&\quad\times\left(\eta\acr_{1,\omega_1}+\sqrt{1-\eta^2}\bcr_{1,\omega_1}\right)\acr_{2,\omega_2}\vac.
\end{align}
\endgroup
It is straightforward to evaluate the derivatives $\ket{\partial_1 \psi}\equiv\ket{\partial_-\psi}$ and $\ket{\partial_2 \psi} \equiv \ket{\partial_+ \psi}$ as
\begin{align}
\ket{\partial_\pm \psi} &= \frac{1}{2\ii}\int_{\R^2} \d\omega_1\d\omega_2\ \e^{-\ii t_{\mathrm{tot}}\frac{\omega_1+\omega_2}{2} - \ii \Delta t\frac{\omega_1-\omega_2}{2}}(\omega_1\pm\omega_2)\notag\\
&\times \bar{\xi}(\omega_1)\bar{\xi}(\omega_2)\left(\eta\acr_{1,\omega_1}+\sqrt{1-\eta^2}\bcr_{1,\omega_1}\right)\acr_{2,\omega_2}\vac
\end{align}
To evaluate the scalar products in Eq.~\eqref{eq:QFIM}, it is useful to first notice that
\begin{align}
&\bra{\mathrm{vac}}\left(\eta\aan_{1,\omega_3}+\sqrt{1-\eta^2}\ban_{1,\omega_3}\right)\left(\eta\acr_{1,\omega_1}+\sqrt{1-\eta^2}\bcr_{1,\omega_1}\right)\notag\\
&\quad\times\aan_{2,\omega_4}\acr_{2,\omega_2}\ket{\mathrm{vac}} = \delta(\omega_1-\omega_3)\delta(\omega_2-\omega_4),
\end{align}
where $\delta(\cdot)$ is the Dirac delta distribution. Now we can easily evaluate the scalar products
\begingroup
\allowdisplaybreaks
\begin{align}
\braket{\partial_- \psi}{\partial_- \psi} &= \frac{1}{4}\int_{\R^2} \d\omega_1\d\omega_2\ (\omega_1-\omega_2)^2\bar{\xi}(\omega_1)^2\bar{\xi}(\omega_2)^2\notag\\
&\quad=\frac{1}{2}\sigma^2 \notag\\
\braket{\partial_+ \psi}{\partial_+ \psi} &= \frac{1}{4}\int_{\R^2} \d\omega_1\d\omega_2\ (\omega_1+\omega_2)^2\bar{\xi}(\omega_1)^2\bar{\xi}(\omega_2)^2\notag\\
&\quad=\frac{1}{2}\left(\sigma^2 + 2\omega_0^2\right) \notag\\
\braket{\partial_- \psi}{\partial_+ \psi} &= \frac{1}{4}\int_{\R^2} \d\omega_1\d\omega_2\ (\omega_1^2-\omega_2^2)\bar{\xi}(\omega_1)^2\bar{\xi}(\omega_2)^2=0 \notag\\
\braket{\psi}{\partial_- \psi} &= \frac{1}{2\ii}\int_{\R^2} \d\omega_1\d\omega_2\ (\omega_1-\omega_2)\bar{\xi}(\omega_1)^2\bar{\xi}(\omega_2)^2 = 0 \notag\\
\braket{\psi}{\partial_+ \psi} &= \frac{1}{2\ii}\int_{\R^2} \d\omega_1\d\omega_2\ (\omega_1+\omega_2)\bar{\xi}(\omega_1)^2\bar{\xi}(\omega_2)^2 \notag\\
&\quad= -\ii\omega_0,
\label{eq:DerivativesQFI}
\end{align}
\endgroup
where $\omega_0$ and $\sigma^2$ are the central frequency and variance of the frequency distribution $\bar{\xi}(\omega)^2$, and finally, by substituting Eq.~\eqref{eq:DerivativesQFI} in Eq.~\eqref{eq:QFIM}, we obtain the QFIM
\begin{equation}
\mathcal{H} = \begin{pmatrix}
2\sigma^2 & 0 \\
0 & 2\sigma^2
\end{pmatrix},
\end{equation}
which is diagonal, and the element $H(\Delta t)$ associated with the delay $\Delta t$ is the value in Eq.~\eqref{eq:QFI} appearing in the Quantum Cramér-Rao bound shown in Eq.~\eqref{eq:CRB} in the main text.

\section{Bunching and coincidence probabilities}
\label{app:ProbCal}
Here, we evaluate the probability in Eq.~\eqref{eq:P} that the two photons are observed with frequencies $\omega_1,\omega_2$ in the same and in different output channels.

The balanced beam splitter, on which the two photons in the state $\ket{\psi}$ in Eq.~\eqref{eq:InitState} impinge, can be described with a $2\times 2$ unitary matrix $\UBS$ of transition amplitudes
\begin{equation}
\UBS = \frac{1}{\sqrt{2}}\begin{pmatrix}
1 & -1\\
1 & 1
\end{pmatrix},
\end{equation}
and it acts on the injected probe through the map $\hUBS \acr_i \hUBS^\dag = \sum_{j=1,2} (\UBS)_{ij}\acr_j$, and equivalently for $\bcr_i$. With reference to Eq.~\eqref{eq:InitState}, the two-photon state $\ket{\psi'}$ at the output of the beam splitter thus reads

\begin{widetext}
\begin{align}
\ket{\psi'}=\hUBS\ket{\psi}&=\frac{1}{2}\int_{\R^2}\dd \omega_1\dd\omega_2\  \xi_1(\omega_1)\xi_2(\omega_2)\left(\eta(\acr_{1,\omega_1}-\acr_{2,\omega_1})+\sqrt{1-\eta^2}(\bcr_{1,\omega_1}-\bcr_{2,\omega_1})\right)(\acr_{1,\omega_2}+\acr_{2,\omega_2})\vac\notag\\
&=\frac{1}{2}\int_{\R^2}\dd \omega_1\dd\omega_2\  \xi_1(\omega_1)\xi_2(\omega_2)\Big(\eta(\acr_{1,\omega_1}\acr_{1,\omega_2}-\acr_{2,\omega_1}\acr_{2,\omega_2}+\acr_{1,\omega_1}\acr_{2,\omega_2}-\acr_{2,\omega_1}\acr_{1,\omega_2})+\notag\\
&\quad+\sqrt{1-\eta^2}(\bcr_{1,\omega_1}\acr_{1,\omega_2}-\bcr_{2,\omega_1}\acr_{2,\omega_2}+\bcr_{1,\omega_1}\acr_{2,\omega_2}-\bcr_{2,\omega_1}\acr_{1,\omega_2})\Big)
\label{eq:OutState1}
\end{align}
\end{widetext}
In order to evaluate the probabilities that two photons with frequencies $\omega$ and $\omega'$ are observed in each configuration (bunching or coincidence), it is convenient to further manipulate \eqref{eq:OutState1}, to more easily take into account the indistinguishability of identical photons, so that

\begingroup
\allowdisplaybreaks
\begin{widetext}
\begin{align}
\ket{\psi'}&=\frac{\eta}{2}\int_{\omega_1<\omega_2}\dd \omega_1\dd\omega_2\ \left(\xi_1(\omega_1)\xi_2(\omega_2)+\xi_1(\omega_2)\xi_2(\omega_1)\right)\left[\acr_{1,\omega_1}\acr_{1,\omega_2}-\acr_{2,\omega_1}\acr_{2,\omega_2}\right]\vac\notag\\
&\quad+\frac{\eta}{2}\int_{\omega_1<\omega_2}\dd \omega_1\dd\omega_2\ \left(\xi_1(\omega_1)\xi_2(\omega_2)-\xi_1(\omega_2)\xi_2(\omega_1)\right)\left[\acr_{1,\omega_1}\acr_{2,\omega_2}-\acr_{2,\omega_1}\acr_{1,\omega_2}\right]\vac+\notag\\
&\quad+\frac{\sqrt{1-\eta^2}}{2}\int_{\R^2}\dd \omega_1\dd\omega_2\  \xi_1(\omega_1)\xi_2(\omega_2)(\bcr_{1,\omega_1}\acr_{1,\omega_2}-\bcr_{2,\omega_1}\acr_{2,\omega_2}+\bcr_{1,\omega_1}\acr_{2,\omega_2}-\bcr_{2,\omega_1}\acr_{1,\omega_2})\vac,
\label{eq:PsiPrime}
\end{align}
\end{widetext}
\endgroup
can be written as a sum of three contributes: the first associated with photons ending up in the same channel and in the mode described by the operator $\hat{a}^\dag_{i,\omega_j}$, the second with the photons ending up in different channels in the mode described by the operator $\hat{a}^\dag_{i,\omega_j}$, and the third is the contribution given when the distinguishability in the inner properties other than time and frequency (e.g. polarisation) is observed at the detectors.
Hence, the probability $p_\mathrm{B}([\omega-\delta\omega/2,\omega+\delta\omega/2],[\omega'-\delta\omega/2,\omega'+\delta\omega/2])$ to observe the two photons together in any of the output channels with frequencies $\omega,\omega'$ within the resolution $\delta\omega$ is given by
\begin{widetext}
\begin{align}
&p_\mathrm{B}([\omega-\delta\omega/2,\omega+\delta\omega/2],[\omega'-\delta\omega/2,\omega'+\delta\omega/2]) =\notag\\
=&\int_{\omega-\frac{\delta\omega}{2}}^{\omega+\frac{\delta\omega}{2}}\int_{\omega'-\frac{\delta\omega}{2}}^{\omega'+\frac{\delta\omega}{2}}\dd w_1\dd w_2\ \sum\limits_{i=1,2}\abs{\cav\aan_{i,w_1}\aan_{i,w_2}\ket{\psi' }}^2+\abs{\cav\aan_{i,w_1}\ban_{i,w_2}\ket{\psi' }}^2+\abs{\cav\ban_{i,w_1}\aan_{i,w_2}\ket{\psi'}}^2=\notag\\
=&\int_{\omega-\frac{\delta\omega}{2}}^{\omega+\frac{\delta\omega}{2}}\int_{\omega'-\frac{\delta\omega}{2}}^{\omega'+\frac{\delta\omega}{2}}\dd w_1\dd w_2\ \left(\frac{\eta^2}{2}\abs{\xi_1(w_1)\xi_2(w_2)+\xi_1(w_2)\xi_2(w_1)}^2+\frac{1-\eta^2}{2}\left(\abs{\xi_1(w_1)\xi_2(w_2)}^2+\abs{\xi_1(w_2)\xi_2(w_1)}^2\right)\right)\notag\\
=&\frac{1}{2}\left(\abs{\xi_1(\omega)\xi_2(\omega')}^2+\abs{\xi_1(\omega')\xi_2(\omega)}^2+2\eta^2\re{\xi_1(\omega)\xi_2(\omega')\xi_1^*(\omega')\xi_2^*(\omega)}\right)\delta\omega\ \delta\omega' \equiv p_\mathrm{B}(\omega,\omega')\delta\omega^2,
\label{eq:CalculationPB}
\end{align}
\end{widetext}
where, in the last step, we assumed that the resolution $\delta\omega$ is small enough so that the variations of the integrand are negligible, which is guaranteed if the conditions in Eq.~\eqref{eq:Resolution} are verified.
Similar considerations can be done for the probability of coincidence $p_\mathrm{C}([\omega-\delta\omega/2,\omega+\delta\omega/2],[\omega'-\delta\omega/2,\omega'+\delta\omega/2])$ for photons ending up in different channels
\begin{widetext}
\begin{align}
&p_\mathrm{C}([\omega-\delta\omega/2,\omega+\delta\omega/2],[\omega'-\delta\omega/2,\omega'+\delta\omega/2])=\notag\\
=&\int_{\omega-\frac{\delta\omega}{2}}^{\omega+\frac{\delta\omega}{2}}\int_{\omega'-\frac{\delta\omega}{2}}^{\omega'+\frac{\delta\omega}{2}}\dd w_1\dd w_2\ \Big(\abs{\cav\aan_{1,w_1}\aan_{2,w_2}\ket{\psi' }}^2+\abs{\cav\aan_{1,w_1}\ban_{2,w_2}\ket{\psi' }}^2+\abs{\cav\ban_{1,w_1}\aan_{2,w_2}\ket{\psi' }}^2\notag\\
&+\abs{\cav\aan_{1,w_2}\aan_{2,w_1}\ket{\psi' }}^2+\abs{\cav\aan_{1,w_2}\ban_{2,w_1}\ket{\psi' }}^2+\abs{\cav\ban_{1,w_2}\aan_{2,w_1}\ket{\psi' }}^2\Big)\notag\\
=&\int_{\omega-\frac{\delta\omega}{2}}^{\omega+\frac{\delta\omega}{2}}\int_{\omega'-\frac{\delta\omega}{2}}^{\omega'+\frac{\delta\omega}{2}}\dd w_1\dd w_2\ \left(\frac{\eta^2}{2}\abs{\xi_1(w_1)\xi_2(w_2)-\xi_1(w_2)\xi_2(w_1)}^2+\frac{1-\eta^2}{2}\left(\abs{\xi_1(w_1)\xi_2(w_2)}^2+\abs{\xi_1(w_2)\xi_2(w_1)}^2\right)\right)\notag\\
=&\frac{1}{2}\left(\abs{\xi_1(\omega)\xi_2(\omega')}^2+\abs{\xi_1(\omega')\xi_2(\omega)}^2-2\eta^2\re{\xi_1(\omega)\xi_2(\omega')\xi_1^*(\omega')\xi_2^*(\omega)}\right)\delta\omega\ \delta\omega' \equiv p_\mathrm{C}(\omega,\omega')\delta\omega^2.
\label{eq:CalculationPC}
\end{align}
\end{widetext}

Let us now suppose that the frequency distributions are of the form $\xi_i(\omega)=\bar{\xi}(\omega)\e^{-\ii\omega t_i}$, with $\bar{\xi}(\omega)$ real and independent on $t_i$, for $i=1,2$, so that, following from Eqs.~\eqref{eq:CalculationPB} and~\eqref{eq:CalculationPC},
\begin{align}
p_\mathrm{B}(\omega,\omega')&=\bar{\xi}(\omega)^2\bar{\xi}(\omega')^2\left(1+\eta^2\cos((\omega-\omega')\Delta t)\right),\notag\\
p_\mathrm{C}(\omega,\omega')&=\bar{\xi}(\omega)^2\bar{\xi}(\omega')^2\left(1-\eta^2\cos((\omega-\omega')\Delta t)\right).
\end{align}
In the case of imperfect detectors detecting a single incoming photon with probability $\gamma^2$, the probability distributions associated with the events of two-photon bunching and two-photon coincidences at the detectors are respectively
\begingroup
\allowdisplaybreaks

\begin{align}
P_\eta^{\mathrm{B}}(\omega,\omega')&=\gamma^2 p_\mathrm{B}(\omega,\omega')\notag\\
&=\gamma^2\bar{\xi}(\omega)^2\bar{\xi}(\omega')^2\left(1+\eta^2\cos((\omega-\omega')\Delta t)\right) \notag\\
P_\eta^{\mathrm{C}}(\omega,\omega')&=\gamma^2p_\mathrm{C}(\omega,\omega')\notag\\
&=\gamma^2 \bar{\xi}(\omega)^2\bar{\xi}(\omega')^2 \left(1-\eta^2\cos((\omega-\omega')\Delta t)\right).
\label{eq:Probabilities}
\end{align}
\endgroup
from which we find the expression in Eq.~\eqref{eq:P}. 
On the other hand, as one may expect, the probabilities 

\begin{align}
P_0&=\gamma^2\notag\\
P_1(\omega)&=\gamma(1-\gamma)\left(\int_\R \dd \omega'\ \left(p^{\mathrm{B}}_\eta(\omega,\omega')+p^{\mathrm{C}}_\eta(\omega,\omega')\right)\right)\notag\\
&=2\gamma(1-\gamma)\bar{\xi}(\omega)^2
\end{align}
of detecting no photons or only a single photon, respectively, do not yield any information on $\Delta t$, and thus do not contribute to the Fisher information.

\section{Fisher information}

\subsection{Arbitrary frequency spectra}
\label{app:FI}

In order to evaluate the Fisher Information in Eq.~\eqref{eq:FisherGen} associated with the estimation of $\Delta t$, we evaluate from the relevant expressions of the probabilities of bunching and coincidence in Eq.~\eqref{eq:Probabilities}

\begin{align}
\frac{\d}{\d\Delta t}&P_\eta^{\mathrm{B}}(\omega,\omega')=-\frac{\d}{\d\Delta t}P_\eta^{\mathrm{C}}(\omega,\omega')\notag\\
&=-\gamma^2\bar{\xi}(\omega)^2\bar{\xi}(\omega')^2\eta^2(\omega-\omega')\sin((\omega-\omega')\Delta t)
\end{align}
and thus, from the definition of Fisher information 
\begin{equation}
F(\Delta t)= \E\left[\left(\frac{\d}{\d\Delta t}\log p(X|\Delta t)\right)^2\right],
\label{eq:FIDef}
\end{equation}
with $X$ denoting the random outcome of the measurement procedure (i.e. frequencies detected and bunching/coincidence detection), we obtain
\begin{align}
F_\eta(\Delta t)&=\int_{\omega<\omega'}\dd\omega\dd\omega'\ \Bigg( \frac{1}{P_\eta^{\mathrm{B}}(\omega,\omega')}\left(\frac{\d}{\d \Delta t}P_\eta^{\mathrm{B}}(\omega,\omega')\right)^2 \notag\\
&\qquad+ \frac{1}{P_\eta^{\mathrm{C}}(\omega,\omega')}\left(\frac{\d}{\d \Delta t}P_\eta^{\mathrm{C}}(\omega,\omega')\right)^2 \Bigg) \notag\\
&=\eta^4\gamma^2\int_{\R^2}\dd\omega\dd\omega'\ \bar{\xi}(\omega)^2\bar{\xi}(\omega')^2(\omega-\omega')^2\notag\\
&\qquad\times\frac{\sin^2((\omega-\omega')\Delta t)}{1-\eta^4\cos^2((\omega-\omega')\Delta t)} = \eta^4\gamma^2\mathcal{I}_\eta(\Delta t),
\label{eq:FisherApp}
\end{align}
which corresponds to Eq.~\eqref{eq:FisherGen}.

\subsection{Regime $\Delta t \gg \tau$}

\label{app:FILargeDelay}

We analyse now the Fisher information $F_\eta(\Delta t)$ in the regime of large time delays $\Delta t \gg \tau$, obtaining the expression in Eq.~\eqref{eq:FisherGenBig}.
We first notice that the function $\zeta_\eta((\omega-\omega')\Delta t)$ in Eq.~\eqref{eq:zeta} oscillates in $\omega-\omega'$, with period $\pi/\Delta t$. 
If the frequency spectrum $\bar{\xi}(\omega)^2$ of the two photons is regular enough (as for Gaussian wave-packets) so that it does not present intrinsic fast fluctuations, we can assume that the rest of the integrand in the Fisher information~\eqref{eq:FisherGen} is essentially constant within the periodicity interval $\pi/\Delta t$ much smaller than the photonic bandwidth $\sigma = 1/(2\tau)$ in the regime $\Delta t \gg \tau$.
It is then possible to substitute $\zeta_\eta((\omega-\omega')\Delta t)$ with its average over its period in Eq.~\eqref{eq:FisherGen}, which reads
\begin{equation}
\frac{\Delta t}{\pi}\int_{[0,\pi/\Delta t]} \d\omega\ \zeta_\eta(\omega\Delta t) = \frac{1-\sqrt{1-\eta^4}}{\eta^4}.
\label{eq:ZetaAvg}
\end{equation}
Replacing $\zeta_\eta((\omega-\omega')\Delta t)$ in \eqref{eq:FisherGen} with the right-hand term in Eq.~\eqref{eq:ZetaAvg}, we obtain
\begin{align}
F_\eta(\Delta t\gg\tau) &= 2 (1-\sqrt{1-\eta^4})\gamma^2 \sigma^2 \notag\\
&= (1-\sqrt{1-\eta^4})F_{\eta=1},
\label{eq:FisherGenBigApp}
\end{align}
as shown in Eq.~\eqref{eq:FisherGenBig}.

\subsection{Gaussian spectra}
\label{app:FIGauss}

Now we obtain the specialised expressions for the Fisher information $\mathcal{F}^\mathrm{G}_\eta(\Delta t)$ shown in Eq.~\eqref{eq:FisherGauss} associated with Gaussian photons, with spectra
\begin{equation}
\bar{\xi}(\omega)^2=\sqrt{\frac{1}{2\pi\sigma^2}}\exp[-\frac{(\omega-\Omega_0)^2}{2\sigma^2}],
\label{eq:GaussianSpectrum}
\end{equation} 
where $\Omega_0$ is the central frequency and $\sigma^2$ the variance. 
By substituting Eq.~\eqref{eq:GaussianSpectrum} in Eq.~\eqref{eq:FisherApp}, we obtain
\begin{align}
F_\eta(\Delta t)=\eta^4\gamma^2\frac{1}{2\pi\sigma^2}\int\d\omega\d\omega' &\e^{-\frac{1}{2\sigma^2}\left((\omega-\Omega_0)^2+(\omega'-\Omega_0)^2\right)}\notag\\
&\times(\omega-\omega')^2\zeta((\omega-\omega')\Delta t).
\end{align}
With a change of variables $\delta=\omega-\omega'$, $s=\omega+\omega'-2\Omega_0$, we get

\begin{align}
F_\eta(\Delta t)&=\eta^4\gamma^2\frac{1}{4\pi\sigma^2}\int\d \delta\, \d s\  \e^{-\frac{1}{4\sigma^2}\left(s^2 + \delta^2\right)}\delta^2\zeta(\delta\Delta t)\notag\\
&=\eta^4\gamma^2\sqrt{\frac{1}{4\pi\sigma^2}}\int\d \delta\, \e^{-\frac{1}{4\sigma^2} \delta^2}\delta^2\zeta(\delta\Delta t),\notag\\
&=\eta^4\gamma^2\sqrt{\frac{\tau^2}{\pi}}\int\d \delta\, \e^{-\tau^2 \delta^2}\delta^2\zeta(\delta\Delta t).
\end{align}
A further change of variable to the dimensionless parameter $\kappa = \delta \tau$, finally yields
\begin{equation}
F_\eta(\Delta t)=\frac{\eta^4\gamma^2}{\tau^2}\sqrt{\frac{1}{\pi}}\int\d \kappa\, \e^{-\kappa^2}\kappa^2\zeta\left(\kappa\frac{\Delta t}{\tau}\right)
\label{eq:FIGaussApp}
\end{equation}
corresponding to the expression for the Fisher information $F^\mathrm{G}_\eta(\Delta t)$ in Eq.~\eqref{eq:FisherGauss}.
From Eq.~\eqref{eq:FIGaussApp}, we can easily obtain the dimensionless quantities
\begin{subequations}
\begin{align}
F_\eta(\Delta t)\Delta t^2=\frac{\Delta t^2}{\tau^2}\eta^4\gamma^2\sqrt{\frac{1}{\pi}}\int\d \kappa\, \e^{-\kappa^2}\kappa^2\zeta\left(\kappa\frac{\Delta t}{\tau}\right)\\
F_\eta(\Delta t)\tau^2=\eta^4\gamma^2\sqrt{\frac{1}{\pi}}\int\d \kappa\, \e^{-\kappa^2}\kappa^2\zeta\left(\kappa\frac{\Delta t}{\tau}\right)
\end{align}
\end{subequations}
which only depend on the dimensionless parameters $\Delta t/\tau$, $\eta$ and $\gamma$.

\subsection{Comparison with non-resolving approach}
\label{app:FINR}
\begin{figure}[t!]
\centering
\includegraphics[width=.95\columnwidth]{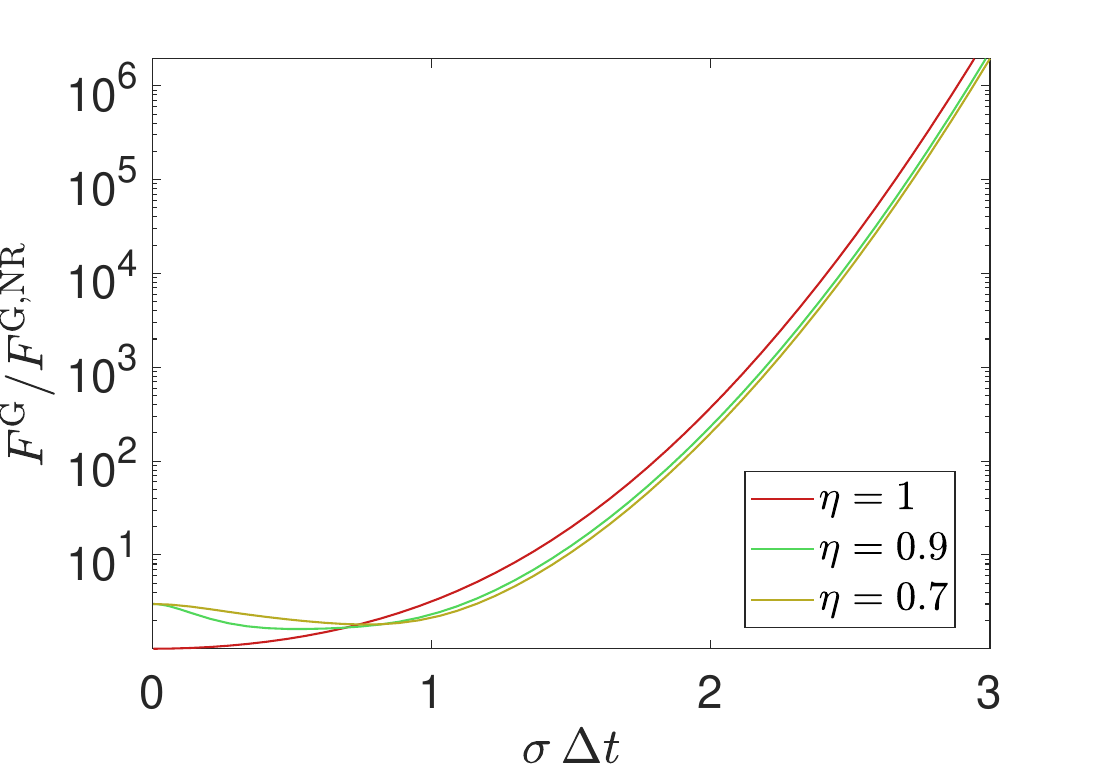}
\caption{Enhancement factor $F^\mathrm{G}/F^\mathrm{G,NR}$ achieved through frequency-resolving detectors with respect to non-resolving techniques, as a function of $\sigma \Delta t \equiv \Delta t/(2\tau)$ and for different values of $\eta$. 
We can see that, for a fixed delay between photons, the enhancement factor increases exponentially with the spectral width $\sigma$, taking values close to $2$ for $\sigma \Delta t \simeq 1$, becoming approximately $10^2$ for $\sigma \Delta t \simeq 2$, and $10^6$ for $\sigma \Delta t \simeq 3$.}
\label{fig:NRatios}
\end{figure}
We will now obtain the expression of the Fisher information in Eq.~\eqref{eq:FisherGaussNoFreq} for non-resolved measurements. If instead the frequencies of the two photons are not observed, the probabilities of bunching and coincidence events can be found by summing the probability in Eq.~\eqref{eq:P} over all the possible frequencies $\omega,\omega'$ obtaining
\begin{subequations}
\label{eq:ProbabilitiesNoFrequency}
\begin{align}
P^\mathrm{B}_\eta = \int_{\omega<\omega'}\dd\omega\dd\omega'\ P^{\mathrm{B}}_\eta(\omega,\omega'), \\
P^\mathrm{C}_\eta = \int_{\omega<\omega'}\dd\omega\dd\omega'\ P^{\mathrm{C}}_\eta(\omega,\omega').
\end{align}
\end{subequations}
\begin{figure*}
\centering
\subfloat{\includegraphics[width=.3\textwidth]{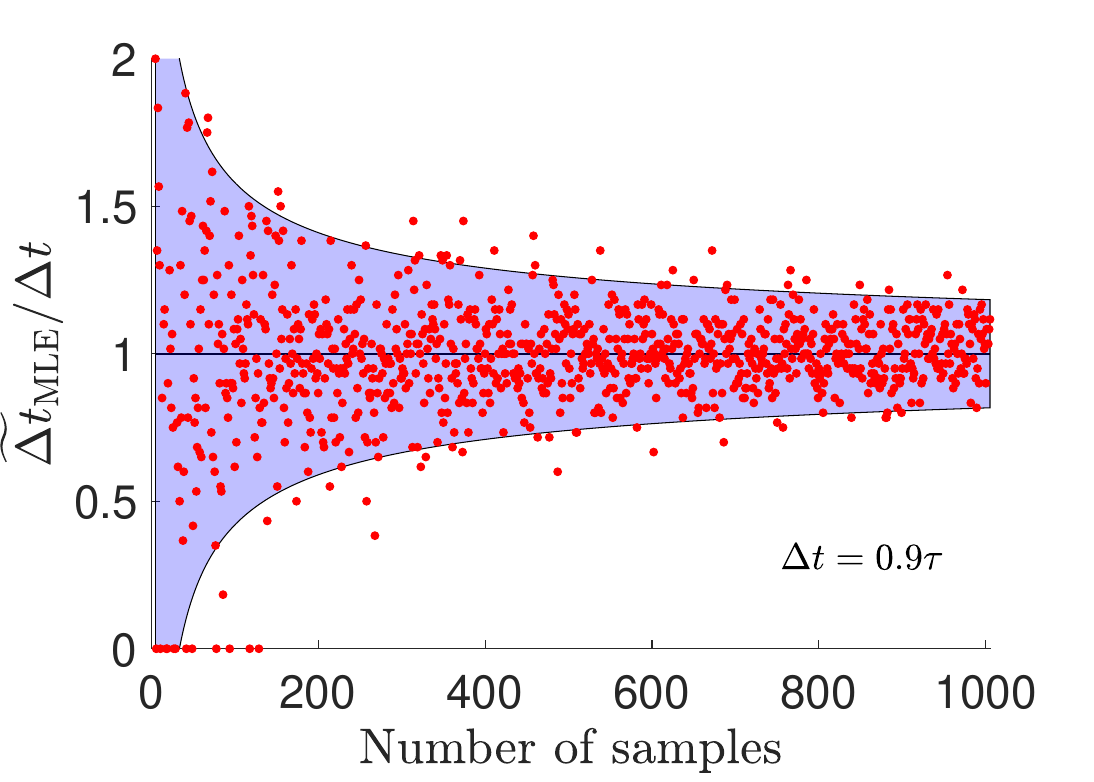}}
\subfloat{\includegraphics[width=.3\textwidth]{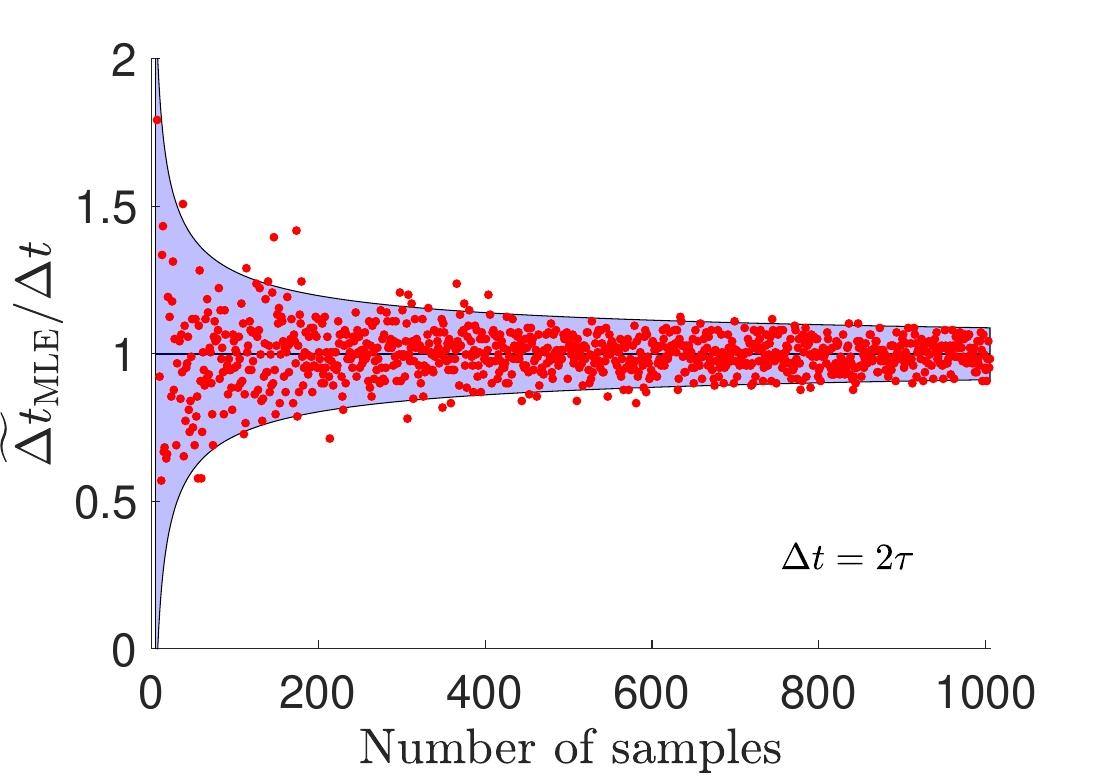}}
\subfloat{\includegraphics[width=.3\textwidth]{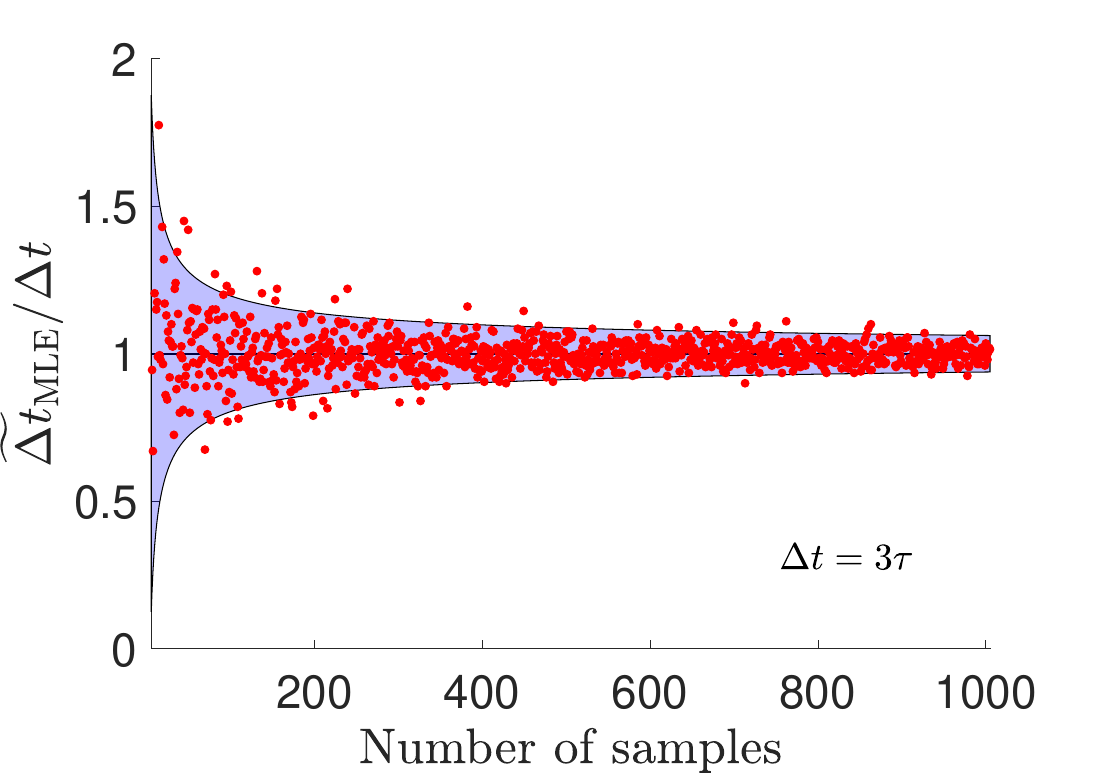}}\\
\subfloat{\includegraphics[width=.3\textwidth]{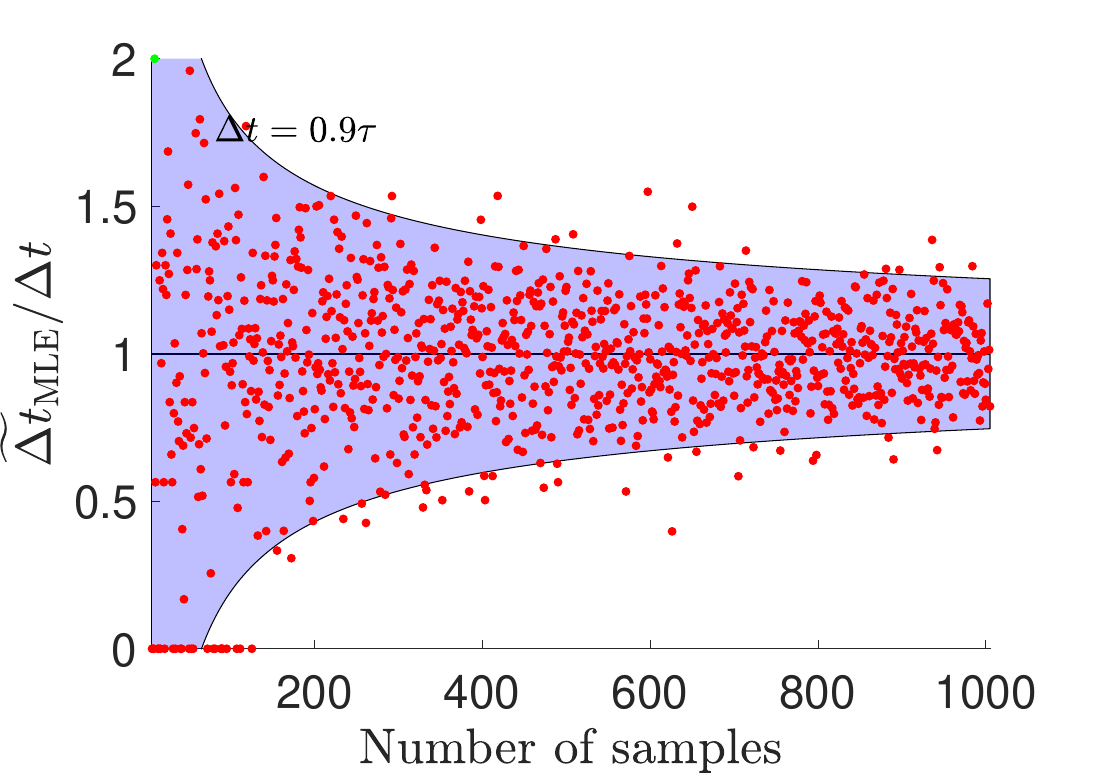}}
\subfloat{\includegraphics[width=.3\textwidth]{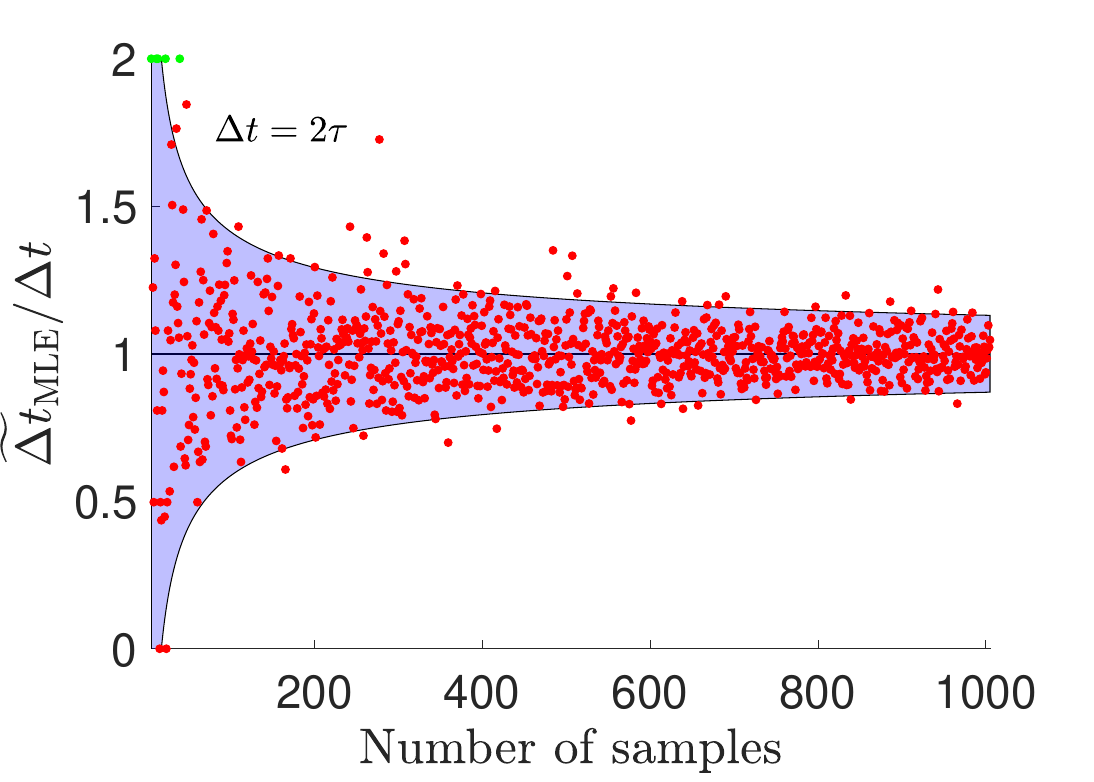}}
\subfloat{\includegraphics[width=.3\textwidth]{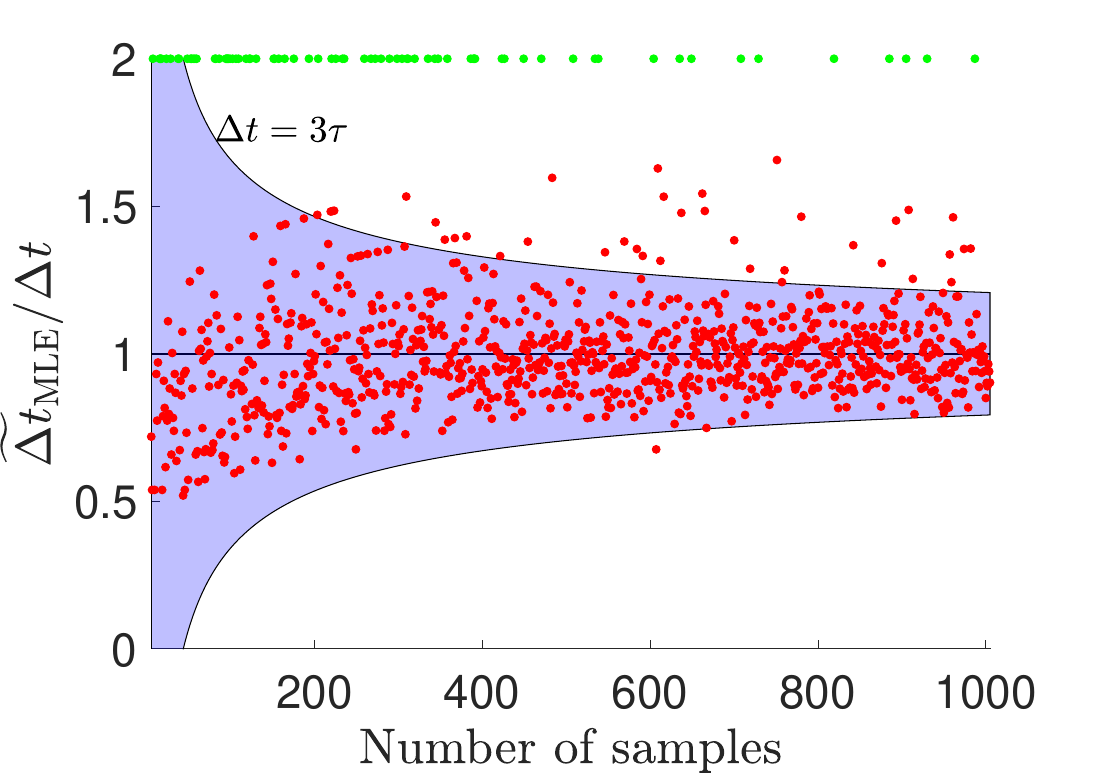}}
\caption{Single outcomes of the estimation of the delay $\Delta t$ for various number of samples, both for the frequency-resolved (upper panels) and non-resolving technique (lower panels) for three different values of the delay, for $\eta\simeq 0.8$.
The area in blue between the two curves in each plot represents a range of two times the Cramèr-Rao bound from the true value of the delay.
We can see how the estimations concentrate much more around $\Delta t$ for the frequency-resolving approach, especially for larger delays, where the non-resolving approach tends to fail.
The green points in the lower panels represent failed estimations of the non-resolving technique: due to statistical fluctuations, more visible with smaller numbers of samples, it is sometimes possible to observe a counting of bunching and coincidence events not compatible with the expressions of the respective probabilities in Eq.~\eqref{eq:ProbNR} for any finite value of $\Delta t$.
These failed estimates are mostly encountered for values of the delay far from the peak of the Fisher information, but also for optimal values of the delay when the indistinguishability $\eta$ at the detectors is small (see \figurename~\ref{fig:MLENumeric} in the main text).
}
\label{fig:MLENumericSingle}
\end{figure*}
Notice how the effect of non resolving the frequencies anticipates the integration over all the frequencies of the photons, which is here performed on the probabilities instead of while evaluating the expectation value of the Fisher information in Eq.~\eqref{eq:FisherGen}.

If we assume that the photons are Gaussian, with spectra given in Eq.~\eqref{eq:GaussianSpectrum}, the probabilities in~\eqref{eq:ProbabilitiesNoFrequency} become
\begin{subequations}

\begin{align}
P^\mathrm{B}_\eta  = \frac{\gamma^2}{2}(1+\eta^2\e^{-\Delta t^2 \sigma^2}),\\
P^\mathrm{C}_\eta  = \frac{\gamma^2}{2}(1-\eta^2\e^{-\Delta t^2 \sigma^2}).
\end{align}
\label{eq:ProbNR}
\end{subequations}

It is straightforward to see, by applying the definition in Eq.~\eqref{eq:FIDef}, that the expression of the Fisher information associated with non-resolving measurements is
\begin{align}
F^{\mathrm{G,NR}}_\eta(\Delta t) &= 4\gamma^2 \frac{\eta^4}{\exp[2\Delta t^2\sigma^2]-\eta^4} \Delta t^2 \sigma^4\notag\\
& \equiv \frac{1}{2} F_{\eta=1} \frac{\eta^4}{\exp[\frac{\Delta t^2}{2\tau^2}]-\eta^4}\frac{\Delta t^2}{\tau^2},
\end{align}
as in Eq.~\eqref{eq:FisherGaussNoFreq} with $F_{\eta=1} =2 \gamma^2\sigma^2 \equiv \gamma^2/(2\tau^2)$.

The advantage of the frequency-resolving technique, compared to non-resolving estimation schemes, is clearly visible in \figurename~\ref{fig:NRatios}, where it is displayed the exponential increase of enhancement factor $F^{\mathrm{G}}(\Delta t)/F^{\mathrm{G,NR}}(\Delta t)$ between the the frequency resolving and non-resolving approaches for increasing values of $\sigma \Delta t$.
It is important to notice that the enhancement factor also represent the ratio $N^{\mathrm{G,NR}}/N^{\mathrm{G}}$ of photon pairs required to reach a given precision with non-resolving and resolving detectors, for schemes saturating the Cramér-Rao bound in Eq.~\eqref{eq:CRB}.
In other words, an enhancement factor of $10^6$, achieved for example for $\sigma \Delta t = 3$, translates into a theoretical reduction of experimental events that need to be observed of a factor $10^6$, with a remarkable impact on the duration of the experiment.

\section{Numerical simulation of an efficient estimator}
\label{app:Num}

We dedicate this section to propose and to analyse an experimental strategy which practically achieves the Cramér–Rao bound in Eq.~\eqref{eq:CRB}.
In particular, this strategy employs the well-known maximum-likelihood estimator, which is reknownedly asymptotically unbiased and efficient, i.e. it saturates the Cramér–Rao bound in the regime of large samples~\cite{Cramer1999,Rohatgi2000}.
We will thus devote our numerical analysis to understand how populated the sample of experimental data must be in order to practically consider this ultimate bound saturated.
We will perform the numerical analysis for the case of Gaussian spectra discussed in the main text.
\begin{figure}[t]
\includegraphics[width=.85\columnwidth]{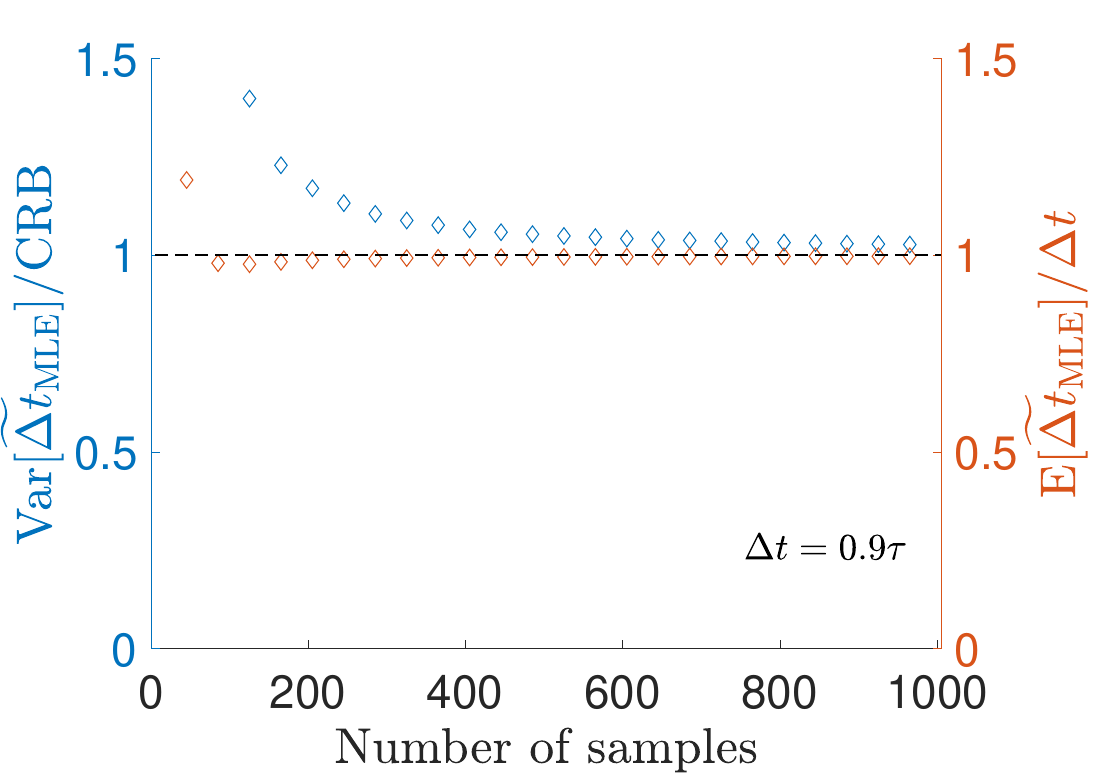}\\
\includegraphics[width=.85\columnwidth]{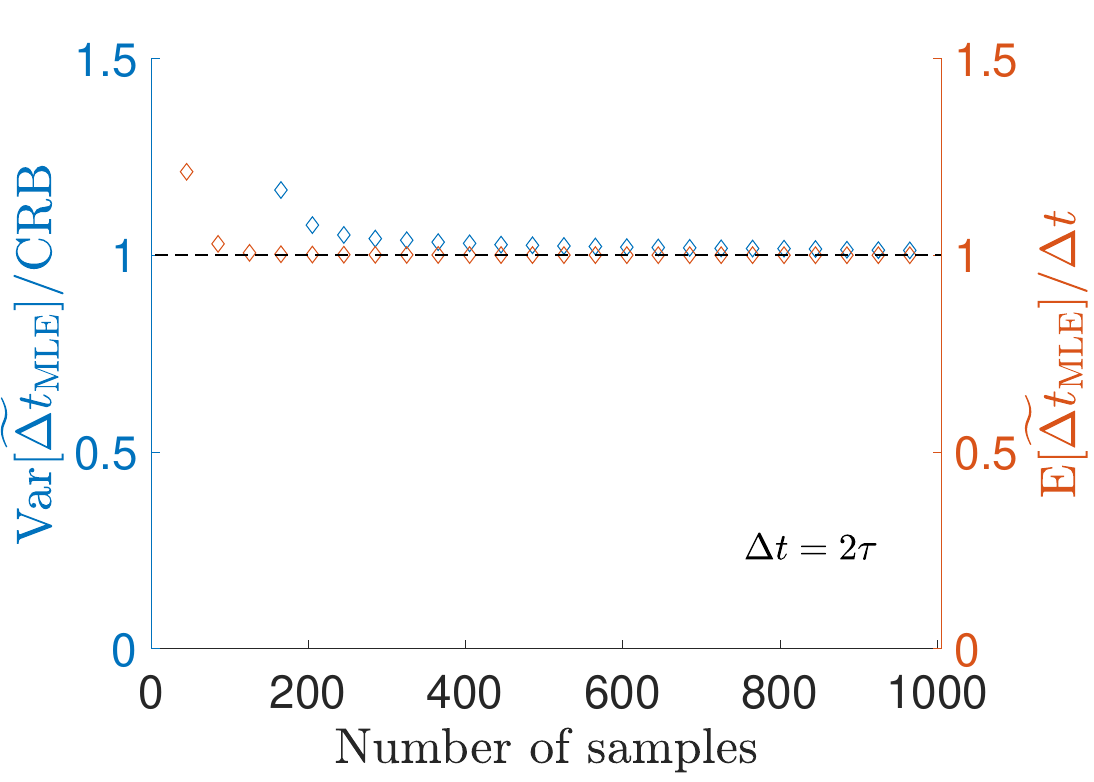}\\
\includegraphics[width=.85\columnwidth]{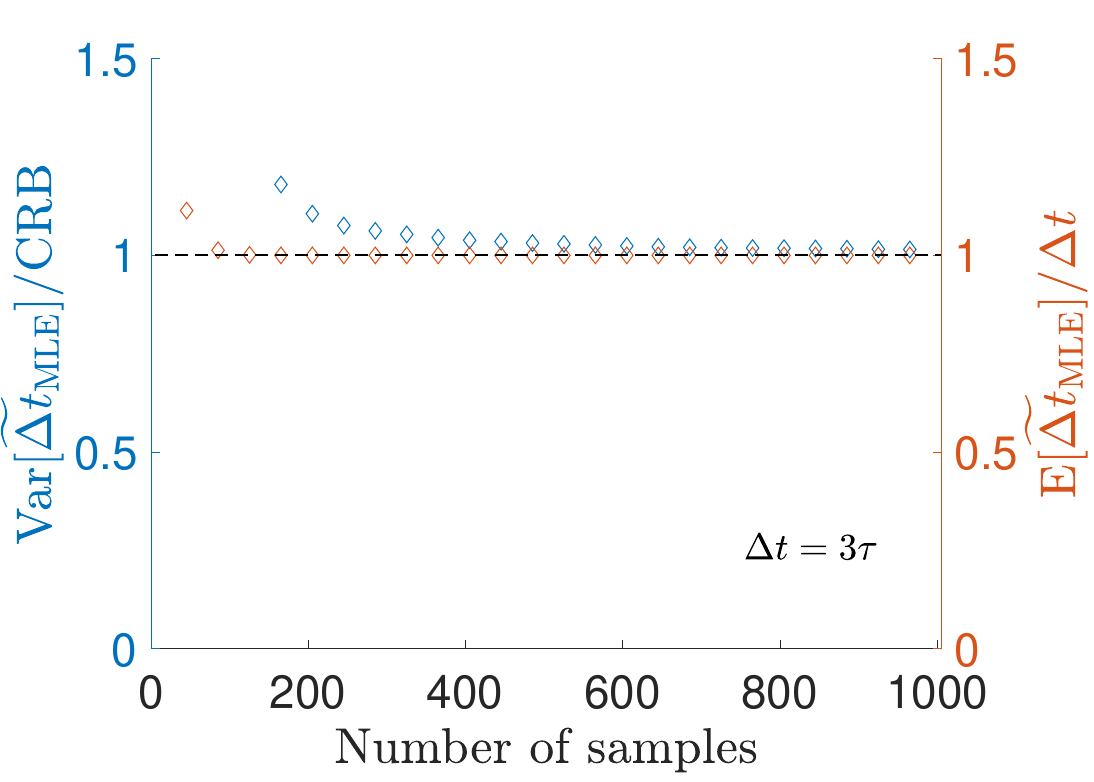}
\caption{Numerical simulation for the estimation of the delay with the frequency-resolving technique have been carried out, and a statistical averages over $10^6$ iterations of the estimation are plotted for the expectation value and the variance of the maximum-likelihood estimator, for $\eta\simeq 0.8$. Both unbiasedness and efficiency are achieved for a numbers of samples smaller than $1000$.}
\end{figure}

Let us imagine to operate the frequency resolving interferometer for a given period of time, in which we manage to observe $N_\gamma \equiv N\gamma^2$ two-photon events.
In doing so, we are already taking into account that a portion $N(1-\gamma^2)$ of events must be discarded since they do not yield any information on the time delay due to losses (only one or no photons are detected).
In fact, we notice from Eqs.~\eqref{eq:CRB} and~\eqref{eq:FisherGen} that the Cramér–Rao bound we evaluate is equivalent to the same bound in a lossless scenario, but with $N\gamma^2$ pairs of photons at our disposal.
We will thus consider from now on a setup without losses, but which employs only a fraction $N_\gamma\equiv N\gamma^2$ of photon pairs.
\begin{figure}[t]
\centering
\subfloat{\includegraphics[width=.9\columnwidth]{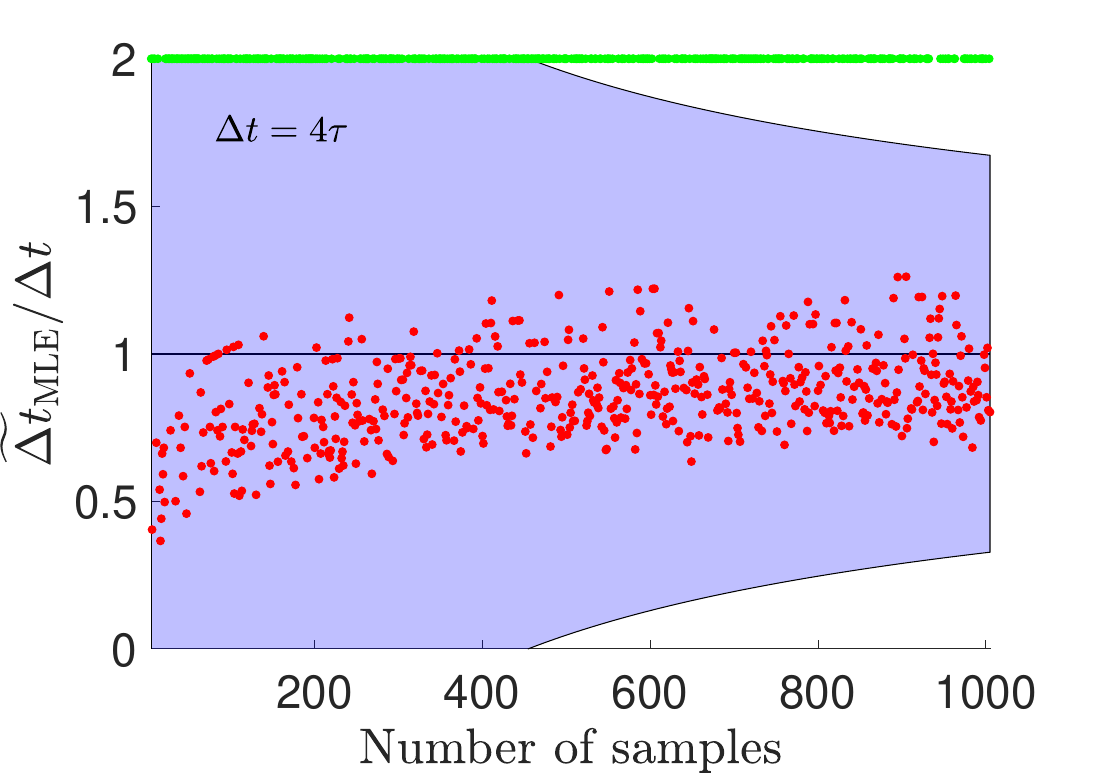}}\\
\subfloat{\includegraphics[width=.9\columnwidth]{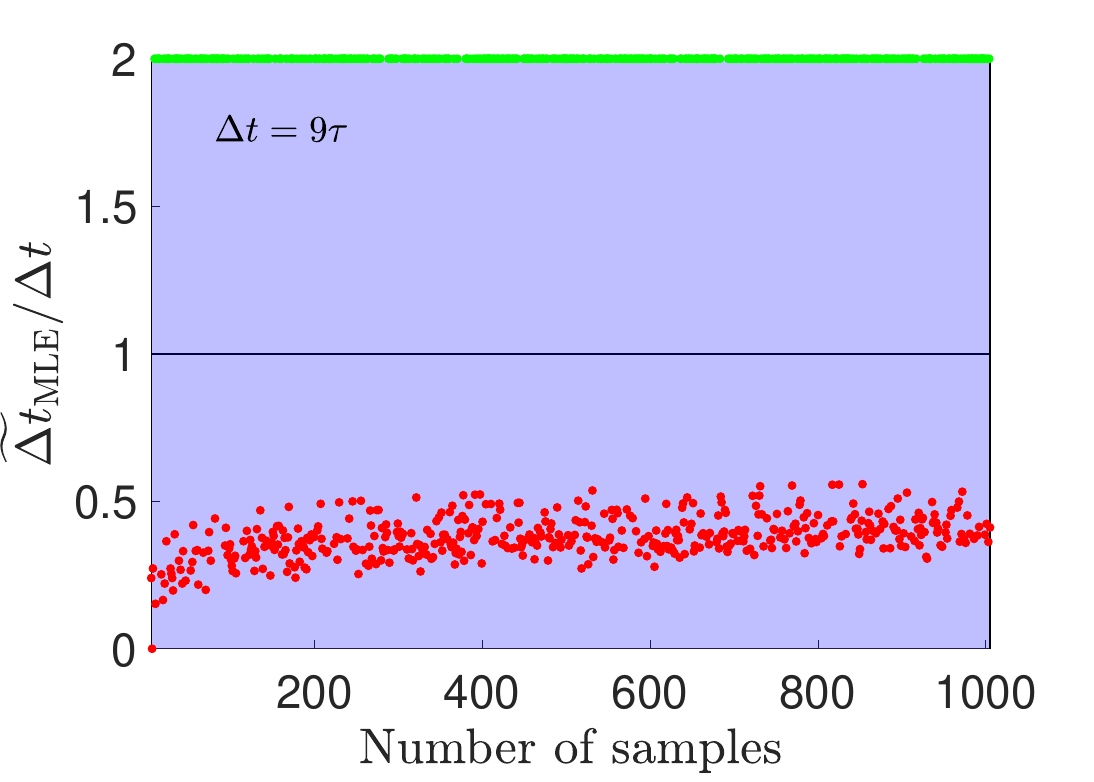}}
\caption{Numerical simulations of the estimation for delays $\Delta t = 4\tau,\,9\tau$ through the non-resolving technique, with $\eta\simeq0.8$. 
In this regime, the Cramèr-Rao bound drastically increases as the Fisher information in Eq.~\eqref{eq:FisherGaussNoFreq} decreases exponentially for increasing $\Delta t$.
Moreover, many iterations failed to yield a finite estimate of the delay due to the statistical fluctuations in the number of coincidence and bunching events observed, as it can be seen from the density of green points, each one representing a failed estimation.
}
\label{fig:MLENumericLargeNR}
\end{figure}

\begin{figure*}[]
\centering
\subfloat{\includegraphics[width=.4\textwidth]{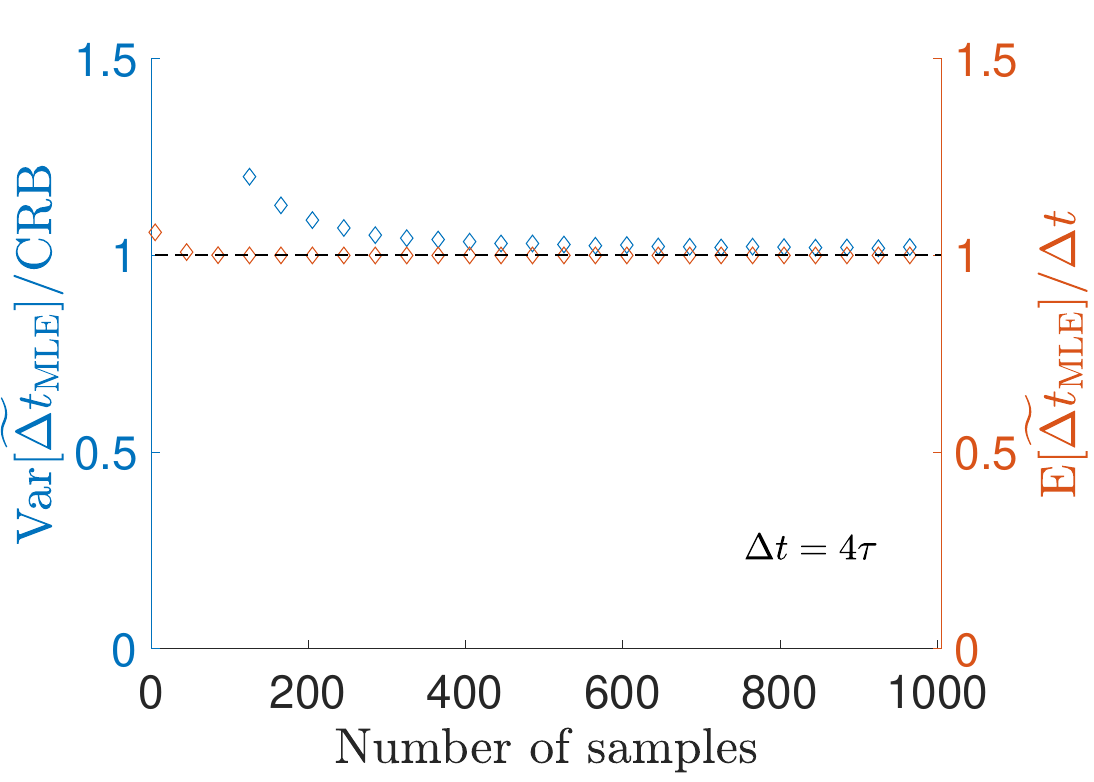}}
\subfloat{\includegraphics[width=.4\textwidth]{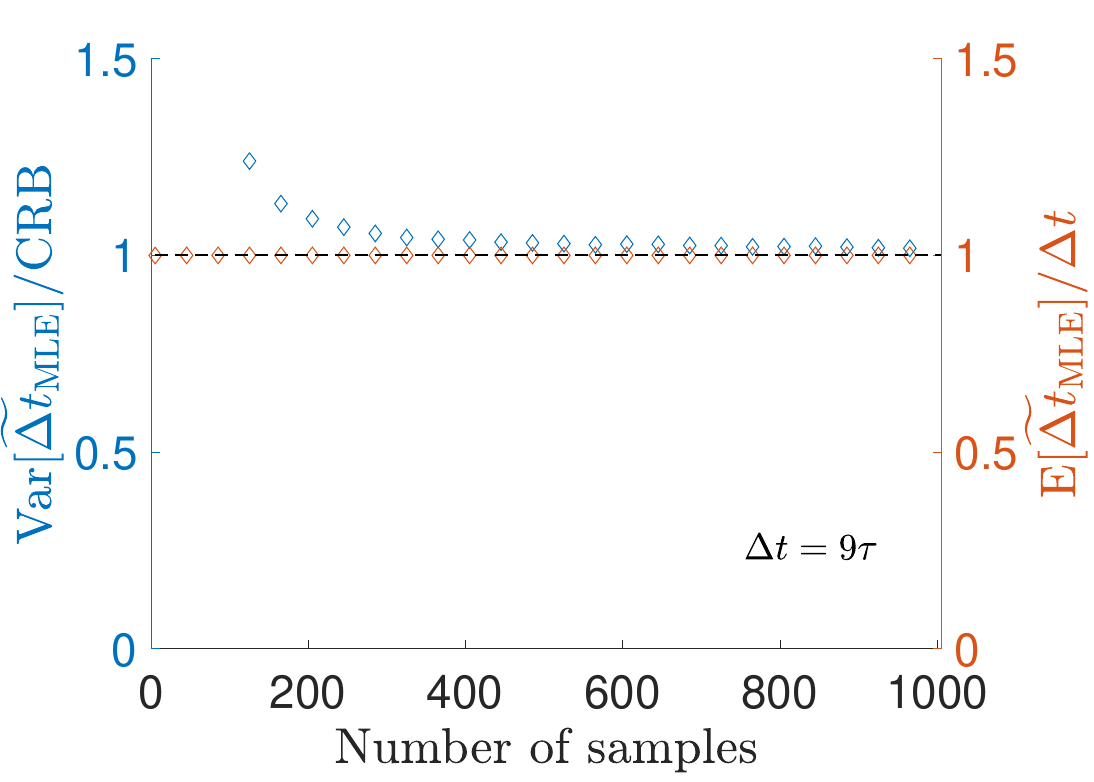}}\\
\subfloat{\includegraphics[width=.4\textwidth]{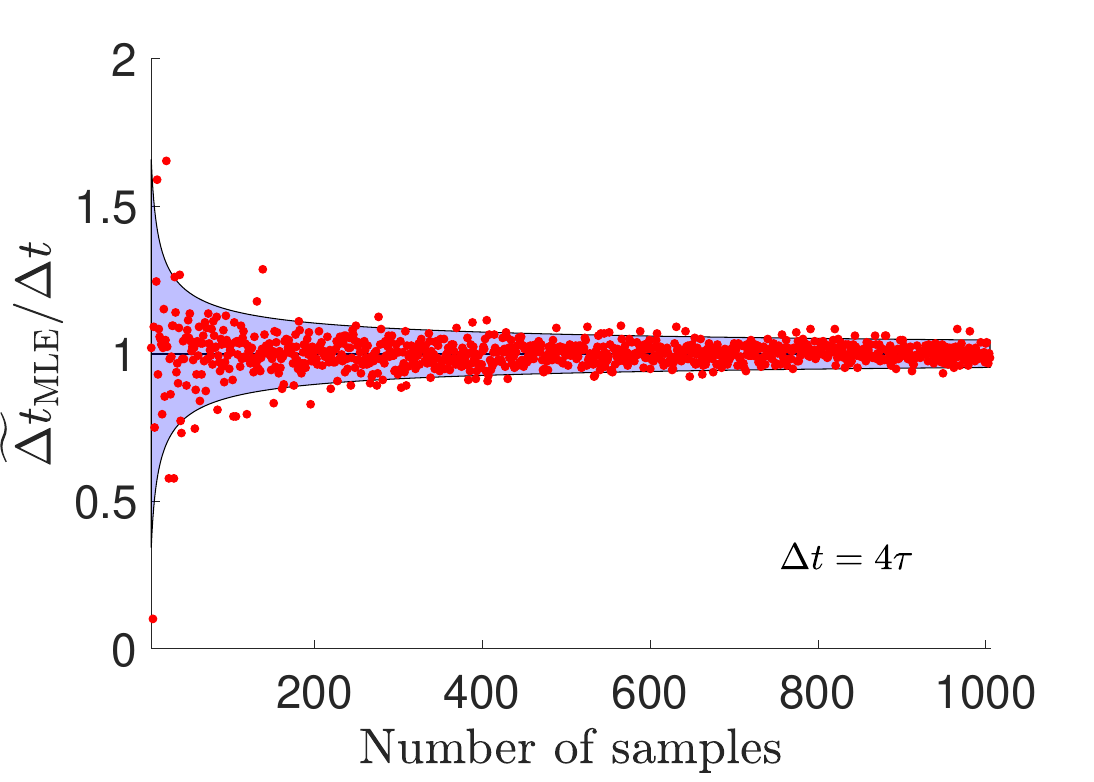}}
\subfloat{\includegraphics[width=.4\textwidth]{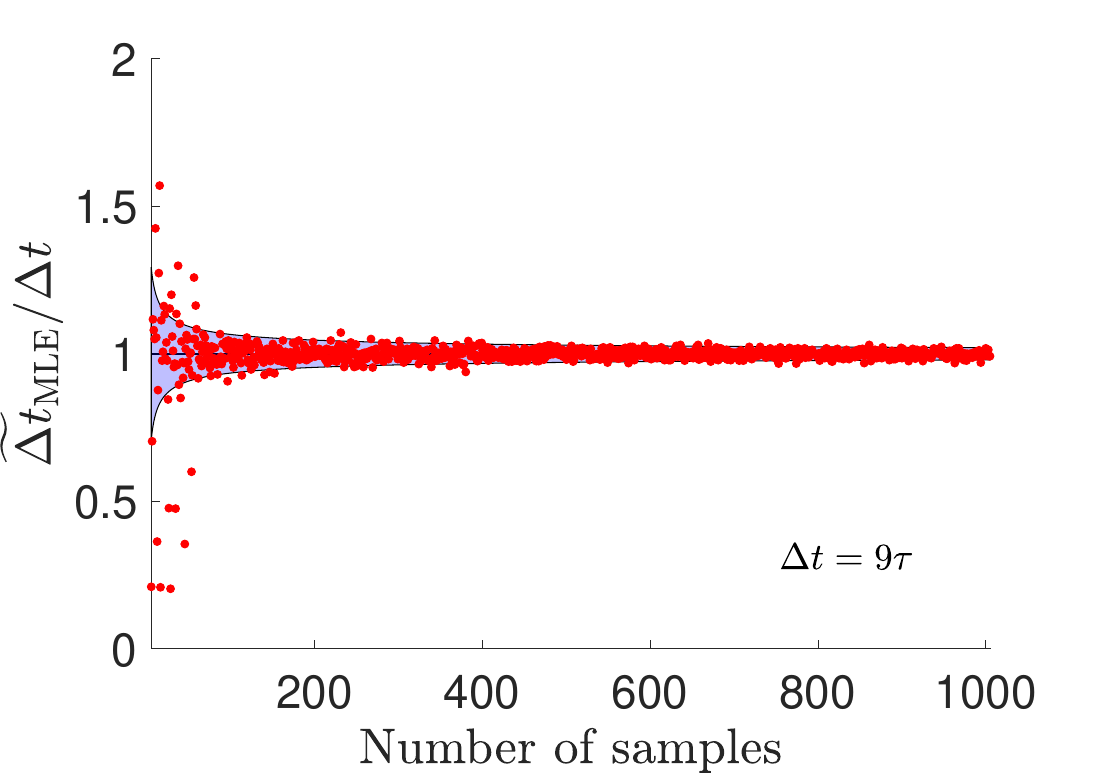}}
\caption{Numerical simulations of the estimation for large delays through our frequency-resolving technique, with $\eta\simeq0.8$. We can see that the unbiasedness and the saturation of the Cramèr-Rao bound are effectively reached within the range of $N_\gamma=1000$ samples observed.
}
\label{fig:MLENumericLargeR}
\end{figure*}

The outcome of the $i$-th observed event $s^{\{i\}}=(\omega_1^{\{i\}},\omega_2^{\{i\}},\mathrm{X}_i)$, with $i=1,\dots,N_\gamma$ will contribute to the experimental sample $S_{N_\gamma} = \{s^{\{i\}}\}_{i=1,\dots,N_\gamma}$ with the joint information regarding the frequencies $\omega_1^{\{i\}}$, $\omega_2^{\{i\}}$ of the two photons, and whether they ended up in the same output channel of the beam-splitter or not, i.e. $\mathrm{X}_i = \mathrm{C}$ for coincidence or $\mathrm{X}_i = \mathrm{B}$ for bunching photons.
Assuming that the resolution $\delta\omega$ of the detectors satisfies conditions~\eqref{eq:Resolution}, and that their working range practically spans over the whole photonic spectra, each outcome of the experiment will be exactly generated according to the probability distribution given by Eq.~\eqref{eq:P} with $\gamma=1$, namely
\begin{equation}
P(s^{\{i\}},\Delta t)\delta\omega^2= P_\eta^{\mathrm{X}_i}(\omega_1^{\{i\}},\omega_2^{\{i\}})\delta\omega^2,
\end{equation}
where we emphasized the dependency of this probability on the unknown delay $\Delta t$.

Since each repetition of the experiment is performed independently from the others, the probability $P(S_{N_\gamma})$ associated with the whole sample $S_{N_\gamma}$ is given by the product 
\begin{equation}
P(S_{N_\gamma},\Delta t)\delta\omega^{2N_\gamma} = \prod_{i=1}^{N_\gamma}P(s^{\{i\}},\Delta t)\delta\omega^2
\label{eq:ProbabilitySample}
\end{equation}
of the single event probabilities.
The expression in Eq.~\eqref{eq:ProbabilitySample} can be thought as a function $\mathcal{L}_\mathrm{R}(\Delta t|S_{N_\gamma})\equiv P(S_{N_\gamma},\Delta t)$ of $\Delta t$, typically called likelihood function, yielding the probability (up to a factor $\delta\omega^{2N_\gamma}$) that the observed sample $S_{N_\gamma}$ was randomly generated according to the value $\Delta t$ of the delay.
Noticeably, for every possible value $\Delta t$, the likelihood $\mathcal{L}_\mathrm{R}(\Delta t| S_{N_\gamma})$ is a number which can be easily evaluated through Eq.~\eqref{eq:ProbabilitySample} once the data sample $S_{N_\gamma}$ has been obtained.
\begin{figure}[b]
\centering
\includegraphics[width=.95\columnwidth]{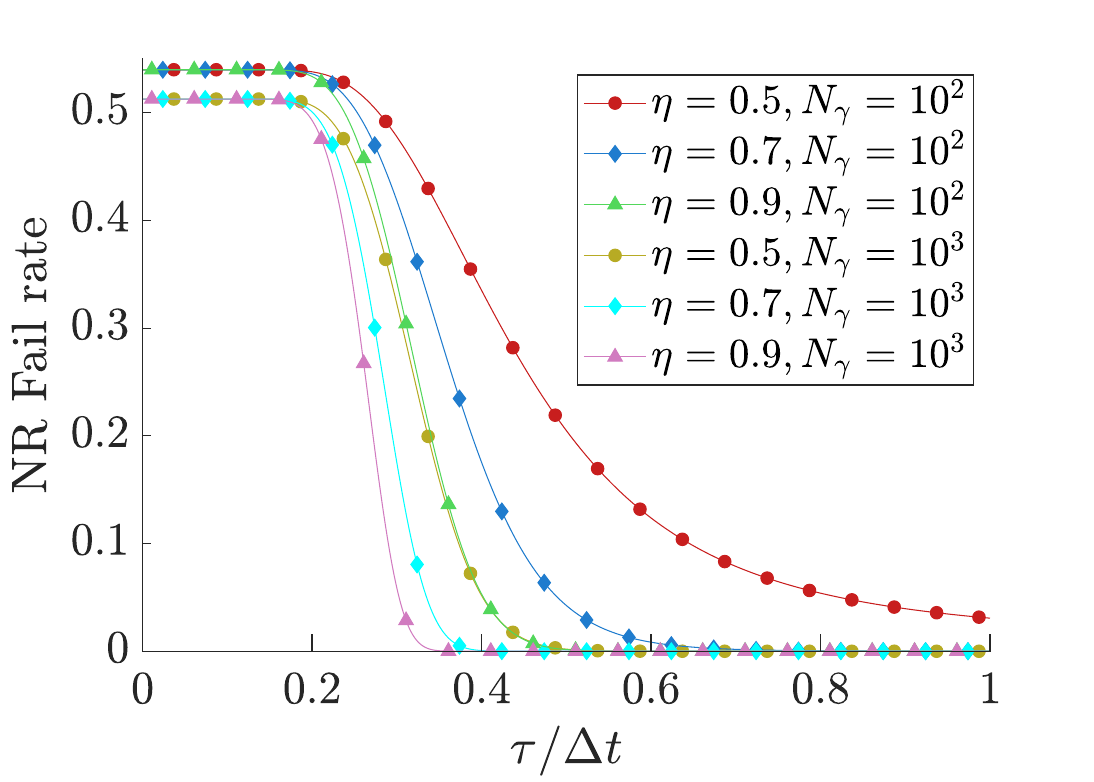}
\caption{Plots of $P_{\mathrm{fail}}$ in Eq.~\eqref{eq:Pfail} as function of $\tau/\Delta t$. Differently from our frequency-resolving technique, the fail rate of the non-resolving approach drastically increases either for smaller values of $\tau/\Delta t$, or for small values of $\eta$ for any value of $\tau/\Delta t$, including the optimal scenarios $\Delta t\simeq 2\tau$ (see \figurename~\ref{fig:FisherGauss}).
}
\label{fig:FailPlot}
\end{figure}

The strategy we propose then makes use of the maximum-likelihood estimator $\widetilde{\Delta t}_{\mathrm{MLE}}$ to retrieve an estimate of the delay true unknown value, which is thus given by the quantity that maximises the likelihood function, i.e. it is implicitly defined by
\begin{equation}
\mathcal{L}_\mathrm{R}(\widetilde{\Delta t}_{\mathrm{MLE}}|S_{N_\gamma}) = \sup_{\Delta t}\mathcal{L}_\mathrm{R}(\Delta t|S_{N_\gamma}).
\label{eq:LikelihoodEquation}
\end{equation}
The advantage of employing the maximum-likelihood estimator is that for large numbers of samples, i.e. large $N_\gamma$, it is asymptotically unbiased and efficient, in the sense that, in this regime, its expectation value equals the true value of the unknown delay $\Delta t$, and its variance saturates the Cramér–Rao Bound, i.e. $\E[\widetilde{\Delta t}_{\mathrm{MLE}}] = \Delta t$ and $\Var{\widetilde{\Delta t}_{\mathrm{MLE}}}=(N \mathcal{F}_\eta(\Delta t))^{-1}$.
On the other hand, eq.~\eqref{eq:LikelihoodEquation} cannot be solved analytically in general, e.g. in the case of Gaussian spectra we examined previously, so a numerical approach is usually undertaken to maximise the likelihood function $\mathcal{L}_\mathrm{R}(\Delta t|S_{N_\gamma})$.

Lastly, in \figurename~\ref{fig:MLENumeric} in the main text and in \figurename s~\ref{fig:MLENumericSingle}-\ref{fig:MLENumericLargeNR}, we test unbiasedness and efficiency of the maximum-likelihood estimator for different numbers $N_\gamma$ of the observed samples $S_{N_\gamma}$, comparing the results of our frequency-resolving strategy with a non-resolving one, for example the time-delay estimation based on maximum-likelihood in Ref.~\cite{Lyons2018}.
We see how both the unbiasedness and the efficiency of the maximum-likelihood estimator are reached for a number $N_\gamma \leqslant 1000$ independently of $\Delta t$ for the frequency-resolving technique (see \figurename s~\ref{fig:MLENumericSingle}-\ref{fig:MLENumericLargeR}).
On the other hand, the non-resolving approach struggles to perform the estimation for values of $\Delta t$ far from the peak of the relative Fisher information (see \figurename s~\ref{fig:MLENumericSingle} and~\ref{fig:MLENumericLargeNR}, and \figurename~\ref{fig:MLENumeric} in the main text).
In this case, not only the Cramèr-Rao bound diverges, yielding a low precision, but also the number of failed estimations, due to statistical fluctuations, increases.

Indeed, each attempt to estimate $\Delta t$ with the non-resolving technique has a generally non-vanishing probability to fail, e.g. to yield a non-finite estimate of the delay.
This happens every time the outcome of the experiment, i.e. the number of bunching and coincidence events $N_\mathrm{B}$ and $N_\mathrm{C}$ respectively, are not compatible with the parametrisation in $\Delta t$ of the relative probabilities $P^\mathrm{B}_\eta$ and $P^\mathrm{C}_\eta$ in Eq.~\eqref{eq:ProbNR}.
In fact, the global maximum of the non-resolved likelihood function 
\begin{align}
\mathcal{L}_\mathrm{NR}(\Delta t|S_{N_\gamma})&= (P^\mathrm{B}_\eta)^{N_\mathrm{B}} (P^\mathrm{C}_\eta)^{N_\mathrm{C}}
\label{eq:LikelihoodNR}
\end{align}
is achieved for
\begin{align}
P^\mathrm{C}_\eta=\frac{N_\mathrm{C}}{N_\mathrm{C}+N_\mathrm{B}},\qquad
  P^\mathrm{B}_\eta = \frac{N_\mathrm{B}}{N_\mathrm{C}+N_\mathrm{B}}.
\label{eq:MaxConditionsNR}
\end{align}
Employing the expressions of $P^\mathrm{C}_\eta$ or $P^\mathrm{B}_\eta$ in Eq.~\eqref{eq:ProbNR}, and solving in $\Delta t$ either of the conditions in Eq.~\eqref{eq:MaxConditionsNR}, one obtains an analytical expression of the maximum-likelihood estimator
\begin{equation}
\Delta t_{\mathrm{NR}}(N_{\mathrm{C}},N_{\mathrm{B}}) =\pm\frac{1}{\sigma} \sqrt{\log\left(\eta^2\frac{N_\mathrm{B}+N_\mathrm{C}}{N_\mathrm{B}-N_\mathrm{C}}\right)}.
\label{eq:AnalMLENR}
\end{equation}
However, for certain outcomes $S_{N_\gamma}=\{N_\mathrm{B},N_\mathrm{C}\}$ of the experiment, the conditions~\eqref{eq:MaxConditionsNR} cannot be satisfied for any real, finite value $\Delta t$ of the delay.
For example, if one observes in a given experiment $N_\mathrm{C}>N_\mathrm{B}$, condition~\eqref{eq:MaxConditionsNR} cannot be satisfied for any real $\Delta t$, since $P^\mathrm{B}_\eta\geqslant P^\mathrm{C}_\eta$ for every real, finite value of $\Delta t$ (see Eqs.~\eqref{eq:ProbNR}), and indeed in this scenario the analytical estimator in Eq.~\eqref{eq:AnalMLENR} yields a complex number.
In this case, the real value of $\Delta t$ maximizing the likelihood function in Eq.~\eqref{eq:LikelihoodNR} is the value for which $P^\mathrm{B}_\eta$ is the smallest and $P^\mathrm{C}_\eta$ the largest possible, i.e. $\Delta t = \infty$.
The probability $P_{\mathrm{fail}}$ that an experiment yields a non-finite estimate of the delay can be calculated as the sum, over all the possible values of $N_\mathrm{C}$ satisfying $N_\mathrm{C}\geqslant N_\mathrm{B}$ (i.e. $N_\mathrm{C}\geqslant N_\gamma/2$) of the probabilities $P(N_\mathrm{C})$ that $N_\mathrm{C}$ coincidence events are observed.
These probabilities follow the binomial distribution, thus the overall probability of failure, plotted in \figurename~\ref{fig:FailPlot}, is given by
\begin{widetext}
\begin{align}
P_{\mathrm{fail}}(N_\gamma,\eta) &= \sum_{N_\mathrm{C}=\left\lceil\frac{N_\gamma}{2}\right\rceil}^{N_\gamma} P(N_\mathrm{C}) = \sum_{N_\mathrm{C}=\left\lceil\frac{N_\gamma}{2}\right\rceil}^{N_\gamma}\binom{N_\gamma}{N_\mathrm{C}} \left(\frac{1-\eta^2\e^{-\Delta t\sigma^2}}{2}\right)^{N_\mathrm{C}}\left(\frac{1+\eta^2\e^{-\Delta t\sigma^2}}{2}\right)^{N_\gamma-N_\mathrm{C}},
\label{eq:Pfail}
\end{align}
\end{widetext}
where $\binom{k}{n}=\frac{n!}{k!(n-k)!}$ is the binomial coefficient, while $\lceil{x}\rceil$ denote the closest integer to $x$ rounded up.

\nocite{*}
\bibliography{references}

\end{document}